\documentclass[USenglish,twocolumn]{article}

\usepackage[utf8]{inputenc}%(only for the pdftex engine)
\usepackage[big]{dgruyter}
\usepackage{babel}
\usepackage{microtype}

\usepackage{graphics} % for pdf, bitmapped graphics files
\usepackage{graphicx}

\usepackage{times} % assumes new font selection scheme installed
\usepackage{amsmath} % assumes amsmath package installed
\usepackage{amssymb}  % assumes amsmath package installed
\usepackage{wrapfig}

\usepackage{tikz} % <--- if tikz pictures are used
\usetikzlibrary{arrows}				
\usetikzlibrary{shapes.misc}
\usetikzlibrary{patterns}
\usetikzlibrary{fit}
\usepackage{pgfplots}
\usetikzlibrary{calc}
\usetikzlibrary{decorations.markings}
\pgfplotsset{every tick label/.append style={font=\small}}
\usepgfplotslibrary{fillbetween}
\usetikzlibrary{decorations.pathreplacing}
\usetikzlibrary{calc, shadings, shadows, snakes, shapes, arrows,trees,patterns}
\usetikzlibrary{chains, fit, automata, positioning}
\usetikzlibrary{math}
\usepackage{pgfplots} % <--- if pgfplots is used

\usepackage{algpseudocode}%
\usepackage[ruled,vlined,linesnumbered]{algorithm2e}
\usepackage{epstopdf}
\usepackage{cases}

\usepackage{color}

\def\SW{\textcolor{blue}}

\usepackage{lipsum}% http://ctan.org/pkg/lipsum
\usepackage{verbatim}

\usepackage{amsthm}
\usepackage{siunitx}

\usepackage{mathtools}
\usepackage{subcaption}

\newtheorem{theorem}{\bf Theorem}

\newtheorem{Algorithm}[theorem]{\bf Algorithm}

\newcommand{\U}{\mathbb{U}}
\newcommand{\R}{\mathbb{R}}
\newcommand{\N}{\mathbb{N}}
\newcommand{\Z}{\mathbb{Z}}
\newcommand{\setS}{\mathbb{S}}

\colorlet{istorange}{orange}
\colorlet{istgreen}{green!50!black}
\colorlet{istblue}{blue} % use structure theme to change
\colorlet{istred}{red!90!black}

\begin{document}

  \articletype{...}

  \author*[1]{Stefan Wildhagen}
  \author[2]{Frank D\"urr}
  \author[3]{Frank Allg\"ower}
  \runningauthor{S. Wildhagen, F. D\"urr, F. Allg\"ower}
  \affil[1]{Universit\"at Stuttgart, Institut f\"ur Systemtheorie und Regelungstechnik, 70550 Stuttgart}
  \affil[2]{Universit\"at Stuttgart, Institut f\"ur Parallele und Verteilte Systeme, 70550 Stuttgart}
  \affil[3]{Universit\"at Stuttgart, Institut f\"ur Systemtheorie und Regelungstechnik, 70550 Stuttgart}
  \title{Rollout event-triggered control: reconciling event- and time-triggered control}
  \runningtitle{Rollout event-triggered control}
  \subtitle{Rollout kann ereignisbasierte und zeitgesteuerte Regelung vers\"ohnen}
  \abstract{Event-triggered control (ETC) and time-triggered control (TTC), the classical concepts to determine the transmission instants for networked control systems, each come with drawbacks: It is difficult to tune ETC such that a certain bandwidth is respected, whereas TTC cannot adapt the sampling interval to the current state of the control system. In this article, we provide an overview over rollout ETC, a method aimed at reconciling the advantages of ETC and TTC. We unite two variants of rollout ETC under a common framework and present conditions for convergence and compliance with a predefined bandwidth limit. Furthermore, we demonstrate that rollout ETC satisfies a performance bound and that it allows for a very flexible transmission scheduling similar to classical ETC. The mentioned beneficial properties are illustrated through extensive numerical simulations.\\
   \vskip2pt  
  \textbf{Zusammenfassung:}
  Sowohl bei der ereignisbasierten als auch der zeitgesteuerten Regelung (ETC, TTC), den klassischen Konzepten zur Bestimmung der Übertragungszeitpunkte für vernetzte Regelungssysteme, gibt es Nachteile: Die Auslegung von ETC, sodass eine bestimmte Bandbreite eingehalten wird, gestaltet sich oft als schwierig und TTC kann die Abtastrate nicht an den aktuellen Zustand der Regelstrecke anpassen. Dieser Artikel gibt einen Überblick über \emph{rollout ETC}, eine Methode die darauf abzielt, die Vorteile von ETC und TTC zu vereinen. Es werden zwei Varianten von rollout ETC in demselben Theorierahmen vereinigt und Bedingungen für Konvergenz und die Einhaltung einer vorher festgelegten Bandbreite präsentiert. Außerdem wird gezeigt, dass rollout ETC eine Mindestregelgüte garantiert und eine sehr flexible Übertragungsplanung, ähnlich zu klassischem ETC, ermöglicht. Die erwähnten vorteilhaften Eigenschaften werden anhand von umfangreichen numerischen Simulationen veranschaulicht.}
  \keywords{Event-triggered control, Networked Control Systems, Model Predictive Control}
  \classification[PACS]{...}
  \communicated{...}
  \dedication{...}
  \received{...}
  \accepted{...}
  \journalname{...}
  \journalyear{...}
  \journalvolume{..}
  \journalissue{..}
  \startpage{1}
  \aop
  \DOI{...}

\maketitle

\section{Introduction}

The defining characteristic of a networked control system (NCS) compared to a classical control loop is that the dedicated communication links between sensors, controllers and actuators are replaced by a shared, digital and possibly wireless communication network \cite{hespanha2007survey}. NCSs have enjoyed an increasing prevalence in recent years, since they are easy to install, highly modular and cheap to maintain. However, the communication network in the control loop also brings additional challenges for control: quantization, transmission delays, packet dropouts and a limited bandwidth are all undesired effects induced by the network. They have the potential to significantly deteriorate control performance and may even lead to instability. Especially when the network is congested, i.e., when the clients request a higher bandwidth than the network can provide, delays and the rate of packet dropouts increase significantly \cite{zhang2012network}.

To address the issue of network congestion, a network manager can assign so-called traffic specifications (TSs) to each of the clients which communicate over the network. TSs restrict the time instances at which a transmission is possible, which allows the network manager to encode certain requirements on the transmission traffic, e.g., a certain bandwidth limit. In return, TSs guarantee that if transmissions are triggered accordingly, congestion in the network is avoided which results in low delays and dropout probabilities, and thereby act as a ``contract'' between network and client \cite{Tanenbaum11}. All relevant types of TSs admit a baseline periodic transmission pattern at a so-called sustainable rate, such that this parameter represents the assigned bandwidth.

In case the network clients are control loops, it is then necessary to design, in addition to the control law, a transmission scheduling strategy which respects the TS. Clearly, classical time-triggered control (TTC) can be easily adjusted such that a certain TS is fulfilled. However, it is inflexible and cannot adjust its sampling interval to the current operating conditions of the control system. For this reason, advanced sampling concepts such as event-triggered control (ETC) \cite{Heemels12} have emerged in recent years. The basic idea behind ETC is that an updated control input is only transmitted when the system requires it, thereby reducing communication compared to traditional TTC. ETC approaches offer a highly flexible transmission scheduling and typically only request transmissions when it is indeed necessary to satisfy control goals (e.g., when disturbances act on the system or when far from a desired setpoint), and dispense with transmissions whenever possible (e.g., when close to the set point). However, it was noted that it is in fact hard to predict the transmission traffic pattern generated by ETC \cite{postoyan2019inter}. Only recently, there have been efforts to characterize ETC traffic patterns a priori, but the results obtained so far apply only to certain system classes and special event-triggering rules \cite{gleizer2020trafficmodels} and/or are computationally demanding \cite{kolarijani2016formal}. Thus, it is a great challenge to design ETC such that it respects a predefined TSs, and often requires many iterations between numerical simulations or formal traffic analysis and redesign.

An alternative method for control and transmission scheduling, geared towards incorporating a prespecified TS, is rollout ETC. Prior work on this method has considered transmission scheduling for the controller-actuator channel \cite{Antunes14,Gommans17,Peters16} as well as for the sensor-controller channel \cite{Koegel19,Rosenthal18}. In contrast to classical ETC, which typically admits an \emph{explicitly} defined control and trigger law to determine when a transmission should take place, in rollout ETC the control and trigger law is \emph{implicitly} defined through the solution to an optimal control problem (OCP). At periodic sampling time instants, rollout ETC measures the current state of the control system and solves the OCP, which amounts to scheduling future transmissions according to the TS and computing the corresponding control updates so that control performance over a finite prediction horizon is maximized. The transmission decision and control applied in closed loop are then the first part of the optimized transmission schedule and input trajectory. By virtue of this repeated open-loop optimization, explicit constraints on the transmission schedule as well as on the state and input can be considered.

To date, the literature on rollout ETC was mainly concerned with the so-called window-based TS \cite{Antunes14,Gommans17,Koegel19,Rosenthal18}. This type of TS subdivides the timeline into disjoint windows of a certain length and allows a fixed number of transmissions in each one of them. The literature on computer networks offers another type of TS, namely the token bucket TS \cite{Tanenbaum11}, which is based on an analogy to a bucket containing tokens. The current filling level of the bucket determines whether a transmission, which would reduce the bucket level by a certain amount of tokens, is allowed or not. The token bucket TS provides more flexibility for control than the window-based TS, since tokens can be saved in the bucket by triggering few transmissions, and then spent later in a burst of transmissions.

In this article, we provide an overview over recent works which investigated the combination of rollout ETC for the controller-actuator channel and the token bucket TS, with the goal to reconcile the advantages of ETC and TTC \cite{jaumann2020saving,Wildhagen19_2,wildhagen2020robustrollout,Wildhagen20}. By leveraging the flexibility of the token bucket TS, rollout ETC can both realize flexible transmissions in terms of many transmissions when needed and few transmissions when possible, and avoid congestion by adhering to the assigned TS. We unite two different variants of rollout ETC under a common framework: the first using a cyclically shrinking prediction horizon and the second periodically time-varying terminal ingredients. For both variants, we demonstrate how to design the OCP such that convergence of the control system and satisfaction of the token bucket TS in closed loop is guaranteed; we indicate that rollout ETC in the first variant never performs worse than a baseline periodic controller; and lastly, we modify the OCP such that the flexibility of the token bucket TS is leveraged. The theoretical results are accompanied by extensive numerical simulations in order to compare the two variants with each other and with ETC and TTC, and to demonstrate that rollout ETC indeed reconciles the advantages of classical ETC and TTC. Following numerous previous works on ETC \cite{astrom2003lebesque,Heemels13_2,Heemels12,tabuada2007event}, NCSs \cite{hespanha2007survey}, and rollout ETC \cite{Antunes14,Gommans17}, we focus on the case of a zero-order hold (ZOH) actuator in this article.

The remainder of this article is structured as follows. After introducing the setup in Section \ref{sec:preliminaries}, we present the rollout ETC algorithm and the two variants in Section \ref{sec:RETC_scheme}. We discuss the theoretical guarantees of both variants in Section \ref{sec:theor_properties}, and accompany each of the results with a numerical example. Finally, we conclude the article and give an outlook in Section \ref{sec:conclusion}.

We write $\N$ for the set of natural numbers, $\N_0\coloneqq \N\cup \{0\}$, $\N_{[a,b]}\coloneqq\N_0\cap[a,b]$ and $\N_{\ge a}\coloneqq\N_0\cap[a,\infty)$, $a,b\in\N_0$. We denote by $I$ the identity matrix and by $0$ the zero matrix of appropriate dimension. We denote by $A\succ0$ $(A\succeq 0)$ a symmetric positive (semi-)definite matrix. For a function $f:\R^n\rightarrow\R^n$, we define the image of the set $S\subseteq\R^n$ as $f(S)\coloneqq\{f(x):x\in S\}$.

\section{Setup} \label{sec:preliminaries}

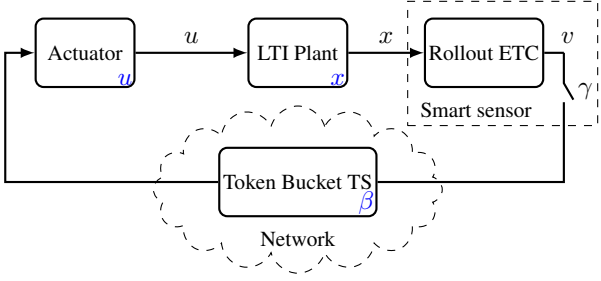
\begin{figure}
	\centering
	\begin{tikzpicture}[>= latex]
	%\tikzmath{\xmax = 7; \ymax =-2; \slength = 0.25;}
	\node [thick,draw,rectangle, inner sep=1.5pt,minimum height = 0.9cm, minimum width = 1.3cm,align = center,rounded corners=3pt] (actuator) at (0.1*7,0) {\small{Actuator}};
	\node[anchor=south east,inner sep=1pt] at (actuator.south east) {\textcolor{istblue}{$u$}};
	\node [thick,draw,rectangle, inner sep=1.5pt,minimum height = 0.9cm, minimum width = 1.3cm,align = center,rounded corners=3pt] (plant) at (0.5*7,0) {\small{LTI Plant}};
	\node[anchor=south east,inner sep=1pt] at (plant.south east) {\textcolor{istblue}{$x$}};
	\node [thick,draw,rectangle, inner sep=1.5pt,minimum height = 0.9cm, minimum width = 1.2cm,align = center,rounded corners=3pt] (ctrl) at (0.85*7,0) {\small{Rollout ETC}};
	\node [thin,dashed,draw,rectangle,minimum height = 1.5cm, minimum width = 2.5cm,align = left] (smartsensor) at (0.885*7,0.07*-2) {\hspace{0.0cm}\\[0.95cm] \hspace{-0.7cm}\small{Smart sensor}};
	\node [thin,dashed,draw,cloud, cloud puffs = 16,minimum height = 2.2cm, minimum width = 3.8cm,align = left] (smartsensor) at (0.5*7,-1.8) {};
	\node (netwname) at (0.5*7,-2.45) {\small{Network}};
	\node [thick,draw,rectangle, inner sep=1.5pt,minimum height = 0.9cm, minimum width = 1.2cm,align = center,rounded corners=3pt] (netw) at (0.5*7,-1.7) {\small{Token Bucket TS}};
	\node[anchor=south east,inner sep=1pt] at (netw.south east) {\textcolor{istblue}{$\beta$}};
	
	\node (upper) at (7,0.2*-2){};
	\node (lower) at (7,0.2*-2-0.25){};
	\node (lower2) at (7+0.25*0.5,0.2*-2 -0.25*0.866){};
	
	\path (0,0) --++ (7,0) node(helpne){} --++ (0,-1.7) node(helpse){} --++ (-7-0.05*7,0) node(helpsw){} --++ (0,1.7) node(helpnw){};
	
	\draw [->,thick] (plant.east) -- (ctrl.west) node[pos = 0.5,above] {$x$};
	\draw [thick] (ctrl.east) -- (helpne.center)node[pos=1.2,above] {$v$};
	\draw [thick] (helpne.center) -- (upper.center) -- (lower2.center) node[pos=0.5,right] {$\gamma$};
	\draw[thick] (lower.center) -- (helpse.center) -- (netw.east);
	\draw[->,thick] (netw.west) -- (helpsw.center) -- (helpnw.center) -- (actuator.west);
	\draw[->,thick] (actuator.east) -- (plant.west) node[pos = 0.5,above] {$u$};
	\end{tikzpicture}
	\caption{Considered configuration of the NCS with token bucket TS.}
	\label{fig:network_tb}
\end{figure}

We consider the NCS configuration depicted in Figure \ref{fig:network_tb}. We will detail each of the components and their interconnection in the following. The controlled plant is described by the discrete-time linear time-invariant (LTI) system
\begin{equation*}
x(k+1) = A x(k) + B u(k), \; x(0) \in \R^n
\end{equation*}
with state $x(k)\in\mathbb{X}\subseteq\R^n$, input $u(k)\in\U\subseteq\R^m$ and time instants $k\in\N_0$, and closed constraint sets $\mathbb{X},\U$ containing the origin. To measure performance, the quadratic cost
\begin{equation*}
x^\top Q x + u^\top R u,
\end{equation*}
is associated with the plant, where $Q,R\succ0$.

As depicted in Figure \ref{fig:network_tb}, updated control inputs $v(k)$ need to be transmitted to the ZOH actuator via a network. The smart sensor, in addition to taking a measurement of the plant in each time step, runs the rollout ETC: based on the measured state, a control update $v(k)$ is calculated and it is determined whether this update should be transmitted via the network to the actuator. The latter is captured in the binary transmission decision
\begin{equation*}
\gamma(k) = \begin{cases}
1 & \text{if a transmission is triggered at time } k \\
0 & \text{if not}
\end{cases}.
\end{equation*}
The control update and transmission decision are computed according to the law
\begin{equation} \label{eq:impl_trig_contr_law}
\begin{bmatrix}	v(k)^\top & \gamma(k)	\end{bmatrix}^\top = \kappa_{\text{RETC}}(x(k),k),
\end{equation}
where $\kappa_{\text{RETC}}:\mathbb{X}\times\N_0\to\U\times\{0,1\}$ is \emph{implicitly} defined by the solution to an OCP to be defined later. Then, if the actuator receives a control update, it applies it to the plant immediately, otherwise it holds the input from the last time step
\begin{equation*}
u(k) = \gamma(k)v(k) + (1-\gamma(k))u(k-1).
\end{equation*}
The previously applied input is saved at the sensor in
\begin{equation*}
	w(k+1) = u(k), \; w(0) = u(-1)\in\U.
\end{equation*}

The token bucket TS is named after an analogy to a bucket containing a variable amount of tokens \cite{Tanenbaum11}. New tokens are added to the bucket at a constant rate of $g\in\N_{\ge 1}$, while triggering a transmission comprises a certain cost $c\in\N_{\ge g}$. The bucket has a maximum capacity $b\in\N_{\ge c}$ and arriving tokens are discarded if the bucket is already full. Hence, the bucket level $\beta$ follows the saturating dynamics
\begin{equation*}
\beta(k+1) = \min\{\beta(k)+g-\gamma(k) c,b\},
\end{equation*}
with $\beta(0)\in\N_{[0,b]}$. A control update can only be transmitted over the network (i.e.,  $\gamma(k)=1$) if the current amount of tokens is sufficient to support the cost of a transmission, which can be described by the constraint $\beta(k)\ge 0$ for all $k\in\N_0$. Hence, the sustainable transmission rate/admissible bandwidth of the token bucket TS is given by $\frac{g}{c}$. In addition, it can be guaranteed a priori that a transmission is possible every $M\coloneqq\lceil\frac{c}{g}\rceil$ time instances. We make the following standing assumptions throughout this article: The parameters $g$, $c$ and $b$ are chosen such that if transmissions satisfy the token bucket TS, congestion in the network is avoided and furthermore, delays and packet dropout probabilities are negligible. The former can be realized in practice by setting the sustainable rate, i.e., $\frac{g}{c}$ at most as high as the bandwidth of the corresponding traffic flow's bottleneck link, and the maximum burst size, i.e., $\lfloor \frac{b}{c} \rfloor$ at most as high as the traffic flow's shortest queue size. In turn, by avoiding congestion in the network, transmission delays and packet dropout probabilities are low and in addition, can be bounded or estimated a priori.

To formulate the OCP and the rollout ETC algorithm in the following, we collect the components of the NCS in the overall state $\xi\coloneqq[x^\top \; w^\top \;  \beta ]^\top$ and input $\pi\coloneqq[ v^\top \; \gamma ]^\top$, the overall dynamics
\begin{equation}
\xi(k+1) = f(\xi(k),\pi(k)),
\label{eq:system_overall}
\end{equation}
with
\begin{equation*}
f(\xi,\pi) \coloneqq \begin{bmatrix}	Ax+B\gamma v + B(1-\gamma)w \\
\gamma v + (1-\gamma)w \\
\min\{\beta+g-\gamma c,b\}
\end{bmatrix},
\end{equation*}
the overall state and input constraints $\xi(k)\in \Xi\coloneqq\mathbb{X}\times\U\times\N_{[0,b]}$, $\pi(k)\in \Pi\coloneqq\U\times\{0,1\}$ and the overall stage cost
\begin{equation} \label{stage_cost}
\ell(\xi,\pi) \coloneqq x^\top Q x + (1-\gamma)w^\top R w + \gamma v^\top R v.
\end{equation}

\section{Rollout ETC \& token bucket TS} \label{sec:RETC_scheme}

The main difference of the token bucket TS compared to the window-based TS is that the filling level of the bucket, which determines when a transmission can be triggered, is a dynamical state itself. When adding the token bucket TS to a control loop, the filling level therefore becomes a component of the NCS's overall state, as we have seen in the previous section. However, convergence of the bucket level is not desired since naturally, it must be decreased from time to time in order to trigger a transmission. Instead, one is typically interested in convergence to the set $\{\xi:(x,w)=0\}$ where the plant state and held input are zero. It is a well-known fact that in receding-horizon methods like rollout ETC or model predictive control (MPC), convergence is not ensured without further ado, and so-called \emph{terminal ingredients} must be added to the underlying OCP in order to give desired guarantees.

In this section, we will show how to combine rollout ETC with the token bucket TS and how to design suitable terminal ingredients. To this end, we will first state the considered rollout ETC algorithm and second, we will introduce the two mentioned variants of rollout ETC, which differ in the particular design of the prediction horizon and terminal ingredients.

\subsection{Rollout ETC algorithm}

Rollout ETC relies on the repeated solution of an open-loop OCP, maximizing the control performance over all possible transmission schedules and control updates. To formulate the OCP, we introduce the predicted overall state and input trajectories at time $k$ as $\xi(\cdot|k)\coloneqq\{\xi(0|k),\ldots,\xi(N(k)|k)\}$ and $\pi(\cdot|k)\coloneqq\{\pi(0|k),\ldots,\pi(N(k)-1|k)\}$, where the prediction horizon $N:\N_0\to\N$ is possibly time-varying. The performance criterion is given by
\begin{equation} \label{eq:MPC_functional}
\begin{aligned}
V(\xi(\cdot|k),\pi(\cdot|k),k)\coloneqq\sum_{i=0}^{N(k)-1} &\ell(\xi(i|k),\pi(i|k)) \\
&+ V_f(\xi(N(k)|k),k),
\end{aligned}
\end{equation}
which sums the overall stage cost \eqref{stage_cost} over the prediction horizon and, in addition, includes a possibly time-varying terminal cost $V_f:\Xi\times\N_0\to\R_{\ge 0}$ penalizing the last predicted state. Given the current state of the overall system $\xi(k)$, the OCP $\mathcal{P}(\xi(k),k)$ solved in each time step is defined by
\begin{subequations} \label{eq:MPC_OP}
	\begin{align}
	&\min_{\xi(\cdot|k),\pi(\cdot|k)} V(\xi(\cdot|k),\pi(\cdot|k),k)  \nonumber \\
	\text{s.t. } &\xi(0|k)=\xi(k) \label{constr_IC} \\
	&\xi(i+1|k) = f(\xi(i|k),\pi(i|k)) \label{constr_dynamics} \\
	&\xi(i|k) \in\Xi, \; \pi(i|k) \in \Pi, \quad \forall i\in\N_{[0,N(k)-1]} \label{constr_state_input}  \\
	&\xi(N(k)|k) \in \Xi_f(k). \label{constr_terminal}
	\end{align}
\end{subequations}
The predicted trajectories are such that they minimize the cost functional \eqref{eq:MPC_functional} over the horizon $N(k)$. The constraints in $\mathcal{P}$ ensure that the predicted trajectories of the plant and the token bucket follow their respective dynamics \eqref{constr_dynamics}, that the constraints of plant state, input and the token bucket TS are fulfilled \eqref{constr_state_input}, and that the initial state in the prediction is the same as the actual measured state \eqref{constr_IC}. Furthermore, a terminal constraint \eqref{constr_terminal} is added to the OCP with a closed and possibly time-varying terminal set $\Xi_f(k)\subseteq\Xi$, for all $k\in\N_0$. We note that due to the integer constraint on $\gamma$, $\mathcal{P}$ is a mixed-integer program and the numerical complexity grows with the prediction horizon on the order of $2^N$. In closed loop, rollout ETC operates according to Algorithm \ref{algo:rollout}.

\begin{algorithm}
	\begin{Algorithm}\label{algo:rollout}
		\normalfont{\textbf{Rollout ETC}}
		\begin{enumerate}
			\item[0.] Set $k=0$.
			\item At time $k$, measure $\xi(k)$, solve the OCP $\mathcal{P}(\xi(k),k)$ \\ and denote the optimizer by $\pi^*(\cdot|k)$.
			\item Apply the first part of the optimal overall input $\pi(k)\coloneqq\pi^*(0|k)$, i.e., if $\gamma^*(0|k)=1$ transmit $v^*(0|k)$ \newline to the actuator, otherwise do not transmit.
			\item Set $k\leftarrow k+1$ and go to Step 1.
		\end{enumerate}
	\end{Algorithm}
\end{algorithm}

After solving $\mathcal{P}(\xi(k),k)$ in Step 1, in Step 2 the predicted overall input is applied to the actual system in terms of $v^*(0|k)$ being transmitted at time $k$ if $\gamma^*(0|k)=1$. This ensures that the overall state and input constraints, i.e., the plant state and input constraints and the token bucket TS, are fulfilled in closed loop as well. Hence, the trigger and control law \eqref{eq:impl_trig_contr_law} is given by
\begin{equation*}
\begin{bmatrix}	v(k)^\top & \gamma(k)	\end{bmatrix}^\top \hspace{-2pt}= \kappa_{\text{RETC}}(x(k),k) \coloneqq \pi^*(0|k) \hspace{-1pt}=\hspace{-1pt} \begin{bmatrix}
v^*(0|k) \\ \gamma^*(0|k)
\end{bmatrix}\hspace{-2pt},
\end{equation*}
where $v^*(0|k)$ and $\gamma^*(0|k)$ depend implicitly on $x(k)$ through the solution to $\mathcal{P}$. We note that in Step 1, in fact the sensor only needs to measure the plant state $x(k)$, whereas $w(k)$ and $\beta(k)$ are internal variables of the controller. This justifies writing $\kappa_{\text{RETC}}$ as dependent on the plant state $x(k)$ and time $k$ only.

\subsection{Design of terminal ingredients \& prediction horizon: two variants}

It is important to note that closed-loop guarantees on constraint satisfaction hold only in case $\mathcal{P}$ admits a solution in each iteration. This property is known as \emph{recursive feasibility} in the literature on receding-horizon control and is not guaranteed without further ado. To ensure recursive feasibility, constraint satisfaction and, in addition, convergence is the purpose of the terminal ingredients, i.e., the terminal constraint \eqref{constr_terminal} and cost $V_f$. As the transmission of a control update is not possible in each time step due to the constraints of the token bucket TS, special care needs to be taken when designing the terminal ingredients. It is in particular not possible to employ time-invariant terminal ingredients \emph{and} a time-invariant prediction horizon at the same time, as typically done in MPC. The literature offers two variants to design the terminal ingredients and the prediction horizon for rollout ETC.

In Variant 1, as followed in \cite{jaumann2020saving,Wildhagen19_2,wildhagen2020robustrollout}, the terminal set as well as the terminal cost are time-invariant
\begin{equation} \label{eq:ing_cyc}
\Xi_f(k) \coloneqq \setS, \quad V_f(\cdot,k) \coloneqq F(\cdot)
\end{equation}

for all $k\in\N_0$, where $\setS\subseteq\Xi$ and $F:\setS\to\R$. At the same time, the prediction horizon is cyclically time-varying
\begin{equation} \label{eq:horizon_cyc}
N(k) \coloneqq \overline{N} - k\bmod M,
\end{equation}
where the cycle length $M$ is the base period of the token bucket TS and $\overline{N}\in\N_{\ge M}$ is the maximum horizon length.

In Variant 1, equivalent to applying Algorithm 1 is to solve $\mathcal{P}$ every $M$ time steps with the full horizon $\overline{N}$, and then to apply the first $M$ parts of the optimal overall input in closed loop. This follows immediately from Bellman's principle of optimality. Such a scheme is less computationally demanding, as the OCP is solved less frequently, but is also less robust w.r.t. disturbances on the plant due to the extended open-loop phase.

Variant 2, as described in \cite{Wildhagen20}, considers a collection of $M$ different terminal ingredients $\{\setS_j,F_j(\cdot)\}_{j=0,\ldots,M-1}$, where $\setS_j\subseteq\Xi$ and $F_j:\setS_j\to\R$ for all $j\in\N_{[0,M-1]}$. Then, the periodically time-varying terminal set and cost are defined by
\begin{equation} \label{eq:ing_per}
\Xi_f(k) \coloneqq \setS_{k\bmod M}, \quad V_f(\cdot,k) \coloneqq F_{k\bmod M}(\cdot)
\end{equation}
for all $k\in\N_0$. The prediction horizon, in contrast, is time-invariant and is given for an $\overline{N}\in\N$ by
\begin{equation} \label{eq:horizon_per}
N(k) \coloneqq \overline{N}.
\end{equation}

For both variants, in the following we will design suitable terminal ingredients to ensure recursive feasibility, constraint satisfaction and convergence.

\subsubsection{Variant 1: cyclic prediction horizon} \label{sec:variant_1}

In the works \cite{Koegel13,Wildhagen19}, it was proven that the desired guarantees hold with time-invariant terminal ingredients and a cyclically time-varying prediction horizon, if there exists a control policy which renders the terminal set $M$-step control invariant \cite[Assumption 2]{Wildhagen19} and decreases the terminal cost over the course of $M$ time steps \cite[Assumption 3]{Wildhagen19}. The idea is now to design a suitable control policy, terminal set and terminal cost for the considered NCS such that these conditions are fulfilled.

Recall that according to the token bucket TS, a feasible schedule is to transmit every $M$ time steps. Inspired by this finding, we consider the following control policy
\begin{equation} \label{eq:terminal_control_policy}
\rho_0(\xi) \coloneqq \begin{bmatrix} K\left[\substack{x\\w}\right] \\ 1 \end{bmatrix}, \quad \rho_j(\xi) \coloneqq \begin{bmatrix} 0 \\ 0 \end{bmatrix}, \; \forall j\in\N_{[1,M-1]},
\end{equation}
where $K\in\R^{m\times(n+m)}$. At $j=0$, the control update $K\left[\substack{x\\w}\right]$ is transmitted, whereas there is no transmission triggered at all other time instances. Defining
\begin{equation*}
\mathcal{A}_1\coloneqq\begin{bmatrix} A & 0 \\ 0 & 0 \end{bmatrix} + \begin{bmatrix} B \\ I \end{bmatrix}K, \quad \mathcal{A}_0\coloneqq\begin{bmatrix} A & B \\ 0 & I \end{bmatrix},
\end{equation*}
we note that $\mathcal{A}_1$ describes the dynamics of $x$ and $w$ under the control $\rho_0$, and $\mathcal{A}_0$ does so under $\rho_j$, $\forall j\in\N_{[1,M-1]}$. Consider a closed set containing the origin $\Z_0\subseteq\mathbb{X}\times\U$, which satisfies
\begin{equation} \label{eq:invariant_set_cyc}
\mathcal{A}_0^j\mathcal{A}_1\Z_0\subseteq\mathbb{X}\times\U, \; \forall j\in\N_{[0,M-2]}, \quad \mathcal{A}_0^{M-1}\mathcal{A}_1\Z_0\subseteq\Z_0,
\end{equation}
i.e., which is $M$-step invariant under the control policy \eqref{eq:terminal_control_policy}. We elaborate in the subsequent subsection on how to construct this set $\Z_0$. Then, if the terminal set is chosen to
\begin{equation} \label{eq:terminal_set_cyc}
\setS \coloneqq \Z_0\times\N_{[c-g,b]},
\end{equation}
the terminal set itself is $M$-step invariant, since 1) the transmission in $\rho_0$ does not cause a violation of the bucket level constraint and 2) the bucket level has recovered to (at least) its initial level after $M$ time steps under the control policy \eqref{eq:terminal_control_policy}. As a result, \eqref{eq:terminal_set_cyc} fulfills \cite[Assumption 2]{Wildhagen19} and is thus a suitable terminal set for the considered NCS.

It remains to design a terminal cost which decreases under application of the control policy \eqref{eq:terminal_control_policy} as required by \cite[Assumption 3]{Wildhagen19}. This can be achieved by the quadratic penalty
\begin{equation} \label{eq:terminal_cost_cyc}
F(\xi)\coloneqq \left[\substack{x\\w}\right]^\top P_0 \left[\substack{x\\w}\right],
\end{equation}
where the weight matrix $P_0\in\R^{(n+m)\times (n+m)}$ satisfies
\begin{align}
&\mathcal{A}_1^\top\mathcal{A}_0^{M-1,\top} P_0 \mathcal{A}_0^{M-1}\mathcal{A}_1 - P_0 + \left[\substack{Q\;0 \\ 0\; 0}\right] + K^\top \left[\substack{R\;0 \\ 0\; 0}\right] K \nonumber \\
&+ \sum_{j=1}^{M-1} \mathcal{A}_1^\top\mathcal{A}_0^{j-1,\top} \left[\substack{Q\;0 \\ 0\; R}\right]\mathcal{A}_0^{j-1}\mathcal{A}_1 \preceq 0. \label{eq:cost_decrease_cyc}
\end{align}
If the pair $(A^M,B_M)$, $B_M\coloneqq\sum_{i=0}^{M-1} A^iB$, is controllable, an admissible choice is the following \cite{Wildhagen19_2}: $P_0\coloneqq\left[\substack{P_x\;0 \\ 0\; 0}\right]$ and $K\coloneqq\left[K_x\;0\right]$, where $P_x$ and $K_x$ are the optimal cost and gain matrix corresponding to the algebraic Riccati equation (ARE) with parameters $A^M,B_M$ and $T_M\coloneqq\sum_{i=0}^{M-1} \mathcal{A}_0^{i,\top} \left[\substack{Q\;0 \\ 0\; R}\right] \mathcal{A}_0^{i}$.

\subsubsection{Variant 2: periodic terminal ingredients} \label{sec:variant_2}

In case of a constant prediction horizon and periodically time-varying terminal ingredients, the works \cite{Lazar18,Wildhagen20} showed that the desired guarantees can be established if there exists a control policy such that the terminal sets are periodically invariant \cite[Assumption 2]{Wildhagen20} and the terminal costs decrease in each time step \cite[Assumption 3]{Wildhagen20}. In the following, we will demonstrate how to design the terminal ingredients for rollout ETC such that these conditions can be met.

We consider the same control policy \eqref{eq:terminal_control_policy} as for Variant 1, but we assume existence of $M$ different closed sets containing the origin $\Z_j\subseteq\mathbb{X}\times\U$, for all $j\in\N_{[0,M-1]}$, which satisfy
\begin{equation} \label{eq:invariant_set_per}
\mathcal{A}_1\Z_0\subseteq\Z_1, \: \mathcal{A}_0\Z_j\subseteq\Z_{j+1}, \: j\in\N_{[1,M-2]}, \: \mathcal{A}_0\Z_{M-1}\subseteq\Z_{0},
\end{equation}
i.e., are periodically invariant. We choose the terminal sets as
\begin{equation} \label{eq:terminal_set_per}
\setS_0 \coloneqq \Z_0\times\N_{[c-g,b]}, \; \setS_j \coloneqq \Z_j\times\N_{[(j-1)g,b]}, \; j\in\N_{[1,M-1]},
\end{equation}
which implies that the terminal sets themselves are periodically invariant (\cite[Lemma 1]{Wildhagen20}). In the practically relevant case of polytopic state and input constraints, constructing sets $\Z_j$ which satisfy \eqref{eq:invariant_set_per} is straightforward as summarized in \cite[Remark 2]{Wildhagen20}: either as polytopes based on \cite{Pluymers05}, or as ellipsoids similar to \cite{Lazar18}. It is easy to see that $\Z_0$ constructed in this way also satisfies \eqref{eq:invariant_set_cyc} and can therefore be used for Variant 1. If there are no plant state and input constraints, i.e., $\mathbb{X}\times\U=\R^{n+m}$, it suffices to choose $\Z_j=\R^{n+m}$.

The terminal costs can be chosen as
\begin{equation} \label{eq:terminal_cost_per}
F_j(\xi)\coloneqq \left[\substack{x\\w}\right]^\top P_j \left[\substack{x\\w}\right],
\end{equation}
with $P_j\in\R^{(n+m)\times (n+m)}$ for all $j\in\N_{[0,M-1]}$. The cost decrease condition \cite[Assumption 3]{Wildhagen20} is met if the $P_j$ satisfy
\begin{equation} \label{eq:cost_decrease_per}
\begin{aligned}
&\mathcal{A}_1^\top P_1 \mathcal{A}_1 - P_0 + \left[\substack{Q\;0 \\ 0\; 0}\right] + K^\top \left[\substack{R\;0 \\ 0\; 0}\right] K \preceq 0. \\
&\mathcal{A}_0^\top P_{(j+1)\bmod M} \mathcal{A}_0 - P_j + \left[\substack{Q\;0 \\ 0\; R}\right] \preceq 0, \; \forall j\in\N_{[1,M-1]}.
\end{aligned}
\end{equation}
An efficient search for suitable $P_j$ and $K$ satisfying \eqref{eq:cost_decrease_per} is possible by transforming it into a linear matrix inequality (LMI), which can be solved efficiently using standard semi-definite program solvers. This yields the following condition: if there exist $\mathcal{X}_j=\mathcal{X}_j^\top\succ 0$ and $\mathcal{Y}$ such that
\begin{align*}
&{\small \begin{bmatrix}
	\mathcal{X}_1 & 0 & 0 & \left[\substack{A\;0 \\ 0\; 0}\right]\mathcal{X}_0 + \left[\substack{B \\ I}\right]\mathcal{Y} \\
	\star & Q^{-1} & 0 & \mathcal{X}_0[\substack{I \\ 0}] \\
	\star & \star & R^{-1} & \mathcal{Y} \\
	\star & \star & \star & \mathcal{X}_0
	\end{bmatrix} } \succeq 0, \\
&{\small \begin{bmatrix}
	\mathcal{X}_{(j+1)\bmod M} & 0 & \mathcal{A}_0 \mathcal{X}_j \\
	\star & \left[\substack{Q^{-1}\;0 \\ 0\; R^{-1}}\right] & \mathcal{X}_j \\
	\star & \star & \mathcal{X}_j \\
	\end{bmatrix}} \succeq 0, \forall j\in\N_{[1,M-1]}
\end{align*}
is satisfied, then $P_j\coloneqq\mathcal{X}_j^{-1}$, $j\in\N_{[0,M-1]}$ and $K\coloneqq\mathcal{Y}\mathcal{X}_0^{-1}$ is an admissible choice .

\section{Closed-loop guarantees} \label{sec:theor_properties}

In this section, we will focus on the closed-loop properties of rollout ETC. At first, we will examine recursive feasibility, constraint satisfaction and convergence. Second, we will give a performance bound for rollout ETC, namely Variant 1 will never perform worse than TTC with periodic transmission pattern at the base period $M$. Under certain circumstances, we can even guarantee a strict performance improvement. Third, we will slightly modify rollout ETC such that it leverages the flexibility of the token bucket TS: it will save transmissions when possible in order to fill up the bucket, such that many tokens are available when many transmissions are required.

\subsection{Recursive feasibility, constraint satisfaction and convergence} \label{sec:convergence}

As already mentioned in the previous section, next to recursive feasibility and constraint satisfaction, our goal is convergence to the set $\{\xi:(x,w)=0\}$. However, due to the transmission decision $\gamma$, the overall stage cost \eqref{stage_cost} is not positive definite w.r.t. this set, which is in contrast to the classical assumptions in receding-horizon control. Naturally, since we want to maximize performance of the plant, the bucket level is not represented in the overall stage cost at all. Such a situation can be analyzed using methods from economic MPC \cite{Faulwasser18}, which is concerned with the case when the cost is some general, not necessarily positive definite, function of state and input. If a certain dissipativity condition holds, it is guaranteed that economic MPC schemes still exhibit convergence to a set where long-term operation w.r.t. the cost function is optimal.

The convergence analysis of rollout ETC relies on recognizing that the set of optimal long-term operation is exactly $\{\xi:(x,w)=0\}$ and the overall system \eqref{eq:system_overall} is strictly dissipative w.r.t. this set. Then, recursive feasibility, constraint satisfaction and convergence can be proven similar as in economic MPC using so-called rotated cost functions. The following result summarizes these guarantees.

\begin{theorem} \label{thm:convergence}
	(\cite[Theorem 1]{Wildhagen19_2},\cite[Theorem 2]{Wildhagen20}) Suppose Algorithm \ref{algo:rollout} is applied either with Variant 1 or Variant 2.
	
	Then, if $\mathcal{P}(\xi(0),0)$ is feasible, the closed loop satisfies:
	\begin{itemize}
		\item $\mathcal{P}(\xi(k),k)$ is feasible at any $k\in\N_0$,
		\item $(\xi(k),\pi(k))\in\Xi\times\Pi$, i.e., the plant state and input constraints and the token bucket TS are fulfilled,
		\item $\xi(k)$ converges to $\{\xi:(x,w)=0\}$ as $k\to\infty$.
	\end{itemize}
	\begin{figure*}%[b]%% over both columns
		\vspace{2pt}
		
		\noindent\makebox[\linewidth]{\rule{\textwidth}{0.4pt}}
	\end{figure*}
\end{theorem}

\begin{figure}
	\begin{subfigure}{0.5\textwidth}
		\centering
		\begin{tikzpicture}

\begin{axis}[%
width=\columnwidth,
height=4.5cm,
xmin=1,
xmax=12,
xlabel={Prediction horizon $\overline{N}$},
ymin=0,
ymax=4000,
ylabel={Infinite-horizon cost},
axis background/.style={fill=white},
legend style={at={(0.03,0.03)}, anchor=south west, legend cell align=left, align=left, draw=white!15!black,legend columns=2,transpose legend}
]

\addplot [color=istorange,mark=star, only marks]
table[row sep=crcr]{%
	6	2773.99605960256\\
	7	1884.35206002380\\
	8	1847.18904311811\\
	9	1849.82283723071\\
	10	1930.18685486225\\
	11	2055.65912084703\\
	12	1836.08905340569\\
};
\addlegendentry{RETC Variant 1}

\addplot [color=istgreen,mark=o, only marks]
table[row sep=crcr]{%
	1 3968.33894943509\\								
	2 2962.55927011994\\
	3 1984.61058828469\\
	4 2049.19462184164\\
	5 2026.30658270358\\
	6 2965.18884143473\\
	7	3095.21624954026\\
	8	1855.67364646355\\
	9	1880.02950332949\\
	10	2029.19697793358\\
	11	2001.10080116751\\
	12	1836.04486785978\\
};
\addlegendentry{RETC Variant 2}

\addplot [color=istblue]
table[row sep=crcr]{%
	0	3.009882122644407e+03\\
	12	3.009882122644407e+03\\
};
\addlegendentry{TTC}

\addplot [color=istred]
table[row sep=crcr]{%
	0	2025.11567911095\\
	12	2025.11567911095\\
};
\addlegendentry{ETC}

\end{axis}
\end{tikzpicture}
		\vspace{-13pt}
		\caption{Cost for rollout ETC with different prediction horizons, TTC and classical ETC.}
		\label{fig:cost_comparison_horizon}
	\end{subfigure}
	\begin{subfigure}{0.5\textwidth}
		\centering
\begin{tikzpicture}

\begin{axis}[%
width=\textwidth,
height=4.8cm,
xmin=0,
xmax=100,
xlabel style={font=\color{white!15!black}},
xlabel={Time $k$},
ymin=0,
ymax=15,
ylabel style={font=\color{white!15!black}},
ylabel={Bucket level $\beta(k)$},
axis background/.style={fill=white},
legend style={at={(0.10,0.97)},anchor=north west,legend cell align=left, align=left, draw=white!15!black,legend columns=2,transpose legend}
]
\addplot[const plot, color=istorange] table[row sep=crcr] {%
0	5\\
1	6\\
2	7\\
3	8\\
4	9\\
5	10\\
6	11\\
7	6\\
8	7\\
9	8\\
10	9\\
11	4\\
12	5\\
13	6\\
14	7\\
15	8\\
16	3\\
17	4\\
18	5\\
19	6\\
20	7\\
21	2\\
22	3\\
23	4\\
24	5\\
25	6\\
26	1\\
27	2\\
28	3\\
29	4\\
30	5\\
31	6\\
32	1\\
33	2\\
34	3\\
35	4\\
36	5\\
37	0\\
38	1\\
39	2\\
40	3\\
41	4\\
42	5\\
43	6\\
44	7\\
45	8\\
46	9\\
47	4\\
48	5\\
49	6\\
50	7\\
51	2\\
52	3\\
53	4\\
54	5\\
55	6\\
56	1\\
57	2\\
58	3\\
59	4\\
60	5\\
61	6\\
62	1\\
63	2\\
64	3\\
65	4\\
66	5\\
67	0\\
68	1\\
69	2\\
70	3\\
71	4\\
72	5\\
73	6\\
74	7\\
75	8\\
76	9\\
77	10\\
78	11\\
79	6\\
80	7\\
81	8\\
82	3\\
83	4\\
84	5\\
85	6\\
86	7\\
87	2\\
88	3\\
89	4\\
90	5\\
91	6\\
92	1\\
93	2\\
94	3\\
95	4\\
96	5\\
97	6\\
98	1\\
99	2\\
100	3\\
101	4\\
};
\addlegendentry{RETC Variant 1}

\addplot[const plot, color=istgreen] table[row sep=crcr] {%
	0	5\\
	1	6\\
	2	7\\
	3	8\\
	4	9\\
	5	10\\
	6	11\\
	7	6\\
	8	7\\
	9	8\\
	10	9\\
	11	4\\
	12	5\\
	13	6\\
	14	7\\
	15	8\\
	16	3\\
	17	4\\
	18	5\\
	19	6\\
	20	7\\
	21	2\\
	22	3\\
	23	4\\
	24	5\\
	25	6\\
	26	1\\
	27	2\\
	28	3\\
	29	4\\
	30	5\\
	31	6\\
	32	1\\
	33	2\\
	34	3\\
	35	4\\
	36	5\\
	37	0\\
	38	1\\
	39	2\\
	40	3\\
	41	4\\
	42	5\\
	43	0\\
	44	1\\
	45	2\\
	46	3\\
	47	4\\
	48	5\\
	49	6\\
	50	7\\
	51	8\\
	52	9\\
	53	4\\
	54	5\\
	55	6\\
	56	7\\
	57	2\\
	58	3\\
	59	4\\
	60	5\\
	61	6\\
	62	7\\
	63	2\\
	64	3\\
	65	4\\
	66	5\\
	67	6\\
	68	1\\
	69	2\\
	70	3\\
	71	4\\
	72	5\\
	73	0\\
	74	1\\
	75	2\\
	76	3\\
	77	4\\
	78	5\\
	79	6\\
	80	7\\
	81	8\\
	82	9\\
	83	4\\
	84	5\\
	85	6\\
	86	1\\
	87	2\\
	88	3\\
	89	4\\
	90	5\\
	91	6\\
	92	1\\
	93	2\\
	94	3\\
	95	4\\
	96	5\\
	97	6\\
	98	1\\
	99	2\\
};
\addlegendentry{RETC Variant 2}

\addplot[const plot, color=istblue] table[row sep=crcr] {%
	0	5\\
	1	0\\
	2	1\\
	3	2\\
	4	3\\
	5	4\\
	6	5\\
	7	0\\
	8	1\\
	9	2\\
	10	3\\
	11	4\\
	12	5\\
	13	0\\
	14	1\\
	15	2\\
	16	3\\
	17	4\\
	18	5\\
	19	0\\
	20	1\\
	21	2\\
	22	3\\
	23	4\\
	24	5\\
	25	0\\
	26	1\\
	27	2\\
	28	3\\
	29	4\\
	30	5\\
	31	0\\
	32	1\\
	33	2\\
	34	3\\
	35	4\\
	36	5\\
	37	0\\
	38	1\\
	39	2\\
	40	3\\
	41	4\\
	42	5\\
	43	0\\
	44	1\\
	45	2\\
	46	3\\
	47	4\\
	48	5\\
	49	0\\
	50	1\\
	51	2\\
	52	3\\
	53	4\\
	54	5\\
	55	0\\
	56	1\\
	57	2\\
	58	3\\
	59	4\\
	60	5\\
	61	0\\
	62	1\\
	63	2\\
	64	3\\
	65	4\\
	66	5\\
	67	0\\
	68	1\\
	69	2\\
	70	3\\
	71	4\\
	72	5\\
	73	0\\
	74	1\\
	75	2\\
	76	3\\
	77	4\\
	78	5\\
	79	0\\
	80	1\\
	81	2\\
	82	3\\
	83	4\\
	84	5\\
	85	0\\
	86	1\\
	87	2\\
	88	3\\
	89	4\\
	90	5\\
	91	0\\
	92	1\\
	93	2\\
	94	3\\
	95	4\\
	96	5\\
	97	0\\
	98	1\\
	99	2\\
	100	3\\
	101	4\\
};
\addlegendentry{TTC}

\addplot[const plot, color=istorange, draw=none, mark=star, mark options={solid, istorange}] table[row sep=crcr] {%
6	11\\
10	9\\
15	8\\
20	7\\
25	6\\
31	6\\
36	5\\
46	9\\
50	7\\
55	6\\
61	6\\
66	5\\
78	11\\
81	8\\
86	7\\
91	6\\
97	6\\
};

\addplot[const plot, color=istgreen, draw=none, mark=o, mark options={solid, istgreen}] table[row sep=crcr] {%
	6	11\\
	10	9\\
	15	8\\
	20	7\\
	25	6\\
	31	6\\
	36	5\\
	42	5\\
	52	9\\
	56	7\\
	62	7\\
	67	6\\
	72	5\\
	82	9\\
	85	6\\
	91	6\\
	97	6\\
};

\addplot[const plot, color=istblue, draw=none, mark=triangle, mark options={solid, istblue}] table[row sep=crcr] {%
0	5\\
6	5\\
12	5\\
18	5\\
24	5\\
30	5\\
36	5\\
42	5\\
48	5\\
54	5\\
60	5\\
66	5\\
72	5\\
78	5\\
84	5\\
90	5\\
96	5\\
};

\end{axis}
\end{tikzpicture}%
		\caption{Evolution of bucket level and transmission time points (indicated by marks) for rollout ETC with $\overline{N}=8$, TTC and classical ETC.}
		\label{fig:bucket_level_horizon}
	\end{subfigure}
	\begin{subfigure}{0.5\textwidth}
		\centering
		\input{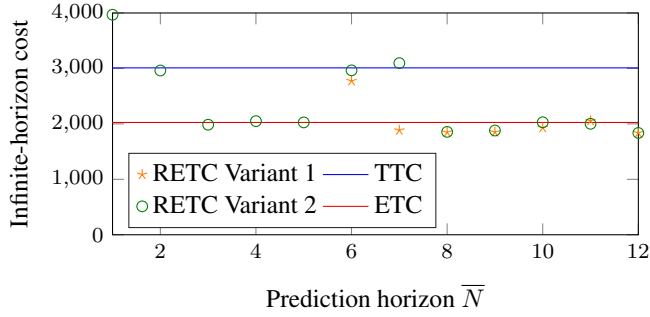}
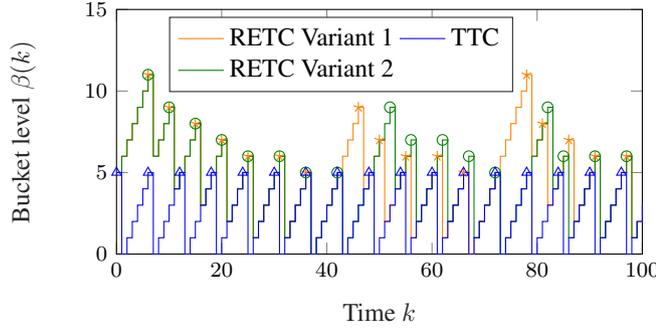
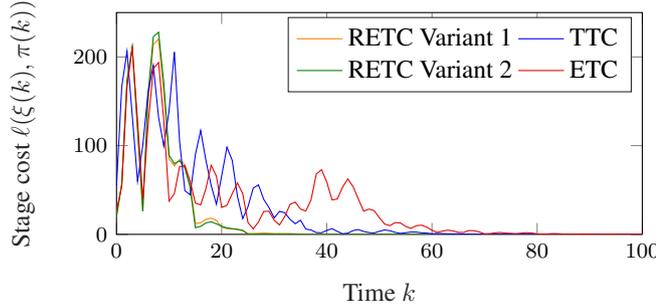
		\caption{Evolution of overall stage cost $\ell$ for rollout ETC with $\overline{N}=8$, TTC and classical ETC.}
		\label{fig:stage_cost}
	\end{subfigure}
	\caption{Cost of rollout ETC, evolution of bucket level and of stage cost.}
\end{figure}

\begin{figure}
	\centering
	\begin{tikzpicture}

\begin{semilogyaxis}[%
width=\columnwidth,
height=4.5cm,
xmin=1,
xmax=12,
xlabel={Prediction horizon $\overline{N}$},
ymin=0,
ymax=7,
ylabel={Computation time in \SI{}{\second}},
axis background/.style={fill=white},
legend style={at={(0.03,0.97)}, anchor=north west, legend cell align=left, align=left, draw=white!15!black}
]

\addplot [color=istorange,mark=star, only marks]
table[row sep=crcr]{%
	6	0.280\\
	7	0.599\\
	8	1.102\\
	9	1.857\\
	10	2.785\\
	11	4.263\\
	12	6.355\\
};
\addlegendentry{RETC Variant 1}

\addplot [color=istgreen,mark=o, only marks]
table[row sep=crcr]{%								
	2 0.039\\
	3 0.067\\
	4 0.110\\
	5 0.182\\
	6	0.280\\
	7	0.599\\
	8	1.102\\
	9	1.857\\
	10	2.785\\
	11	4.263\\
	12	6.355\\
};
\addlegendentry{RETC Variant 2}

\end{semilogyaxis}
\end{tikzpicture}
	\caption{Computation time to solve OCP with different prediction horizons.}
	\label{fig:computation_time_horizon}
\end{figure}
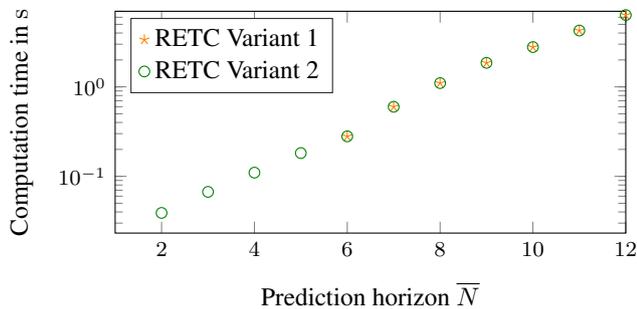

In the following, we will illustrate Theorem \ref{thm:convergence}, compare the two variants and contrast rollout ETC with TTC and classical ETC by means of a numerical example.To this end, we consider the two-mass-spring system from \cite{Antunes14}, discretized with a sampling time of \SI{0.1}{\second}. The system matrices $A$ and $B$ are
\begin{equation*}
\begingroup % keep the change local
\setlength\arraycolsep{2pt}
{\small A = \begin{bmatrix}
	0.9045 & 0.0955 &	0.0968 & 0.0032 \\
	0.0955 &	0.9045 &	0.0032 &	0.0968 \\
	-1.8466 &	1.8466 &	0.9045 &	0.0955 \\
	1.8466 &	-1.8466 &	0.0955 &	0.9045
	\end{bmatrix}\hspace{-3pt}, \:
	B = \begin{bmatrix}
	0.0049 \\
	0.0001 \\
	0.0968 \\
	0.0032
	\end{bmatrix}\hspace{-3pt}.}
\endgroup
\end{equation*}
A token bucket TS has been assigned to the control loop by a network manager with parameters $g=1$, $c=6$ and $b=22$, resulting in a base period of $M=6$ and a sustainable rate/maximum bandwidth of $\frac{g}{c}=\frac{1}{6}\approx 0.1667$. The initial plant state, held input and bucket level are given by $x(0)=[1\;0\;1\;0]^\top$, $w(0)=0$ and $\beta(0)=c-g=5$, making a transmission possible already at $k=0$. The plant state and input are unconstrained, i.e., $\mathbb{X}\times\U = \R^{n+m}$ and the cost matrices are $Q=10I$ and $R=1$. For this system and bandwidth constraint, the optimal time-triggered strategy is to transmit periodically with period $M=6$, and to choose the control gain as discussed in Subsection \ref{sec:variant_1}. With this TTC strategy, the infinite-horizon control cost is given by $x(0)^\top P_x x(0)=3010$.

Figure \ref{fig:cost_comparison_horizon} depicts the infinite-horizon cost of the optimal TTC and of rollout ETC in both variants for different prediction horizons, where the terminal ingredients have been computed using Matlab, YALMIP \cite{YALMIP} and SDPT3 \cite{SDTP3}. The cost of rollout ETC was approximated by simulating the system for a large time until it has converged ($k\in\N_{[0,500]}$) and summing the overall stage cost \eqref{stage_cost} in each time step. We note that for Variant 1, the prediction horizon must be at least $\overline{N}\ge 6=M$, whereas for Variant 2, the prediction horizon can be an arbitrary natural number. From Figure \ref{fig:cost_comparison_horizon}, we see that the cost for rollout ETC is smaller than that of TTC in almost all of the cases, and often even much smaller, although rollout ETC and TTC have the same number of available transmissions. Only in Variant 2 and when $\overline{N}=1$ or $\overline{N}=7$, TTC has better performance. 

In Figure \ref{fig:bucket_level_horizon}, where the bucket level and transmission time points for TTC and rollout ETC with $\overline{N}=8$ are depicted, one can find an explanation for the observed performance gain: while both TTC and rollout ETC fulfill the token bucket TS, the latter is able to flexibly schedule the transmissions in order to maximize control performance, whereas the former has fixed transmission instants. As shown in Figure \ref{fig:stage_cost}, this results in a faster convergence of the stage cost for rollout ETC.

For comparison, we also implemented a classical ETC scheme with the triggering condition from \cite{tabuada2007event}
\begin{equation*}
(x(k')-x(k))^\top(x(k')-x(k))>\sigma x(k)^\top x(k),
\end{equation*}
where $\sigma\in[0,1]$ and $k'$ denotes the last transmission time. The same linear control law as for the optimal TTC is used. We calculated the lowest achievable cost of ETC by densely gridding the parameter $\sigma$ from 0 to 1, simulating for $k\in\N_{[0,500]}$ and comparing the resulting cumulated costs. The ETC with the lowest cost triggered 207 transmissions, which corresponds to an average bandwidth of $0.4132$, thereby significantly violating the sustainable rate of the token bucket TS. From Figure \ref{fig:cost_comparison_horizon}, one can see that the cost of this best classical ETC is still slightly higher than the best achieved cost of rollout ETC, although the latter satisfies the token bucket TS. The best classical ETC which satisfied the average bandwidth constraint had a cost of $2778$, which is only slightly better than TTC.

Lastly, we compare the computation time to solve the OCP in rollout ETC for Variants 1 and 2. To this end, we consider the state and input constraints $\mathbb{X}=[-2,2]^2\times[-5,5]^2$ and $\U = [-12,12]$ and construct polytopic sets $\Z_j$ as discussed in Subsection \ref{sec:variant_2}. In Figure \ref{fig:computation_time_horizon}, the time it took to solve the OCP $\mathcal{P}(\xi(0),0)$ with $\beta(0)=22$ is shown on a logarithmic scale for different prediction horizons. First, we note that if the same prediction horizon is used, the computation times for Variants 1 and 2 are almost equal. Second, the computation times increase exponentially with the prediction horizon $\overline{N}$, as expected. Third, we note that for Variant 1, the prediction horizon must be greater than $M$, which is fixed by the network manager and cannot be altered. This places a lower bound on the computational complexity of Variant 1. A great advantage of Variant 2 is that the prediction horizon, and thereby the computational complexity, can be adjusted independently of $M$, such that much lower computation times can be achieved. This greatly facilitates meeting real-time requirements of the control loop, which are typically of great relevance in practice.

\subsection{Performance} \label{sec:performance}

In the previous numerical example, we have seen that rollout ETC typically has a control performance at least as high as the optimal TTC. In this subsection, we demonstrate that in case of Variant 1, one can even give a guarantee for this.

\begin{theorem} \label{thm:performance_bound}
	(\cite[Theorem 3]{Wildhagen19_2}) Suppose Algorithm \ref{algo:rollout} is applied with Variant 1, $\mathbb{X}\times\U=\R^{n+m}$, $\overline{N} = rM$, $r\in\N$ and $\beta(0)\ge c-g$. Then, the closed loop satisfies
	\begin{equation*}
	\sum_{k=0}^{\infty} \ell(\xi(k),\pi(k)) \le x(0)^\top P_x x(0).
	\end{equation*}
\end{theorem}

According to Theorem \ref{thm:performance_bound}, the infinite-horizon control cost of rollout ETC is less than or equal to that of the optimal TTC fulfilling the TS. The proof relies on recognizing that since the periodic schedule is feasible in the OCP and due to the choice of terminal cost and prediction horizon, the optimal cost \eqref{eq:MPC_functional} is bounded by $x(k)^\top P_x x(k)$. The following result establishes sufficient conditions for a \emph{strict} improvement over TTC.

\begin{theorem} \label{thm:strict_performance_bound}
	(\cite[Theorem 4]{Wildhagen19_2}) Suppose Algorithm \ref{algo:rollout} is applied with Variant 1, the conditions of Theorem \ref{thm:performance_bound} hold, and $\beta(0)\ge 2c-g$. Then, the closed loop satisfies
	\begin{equation*}
	\sum_{k=0}^{\infty} \ell(\xi(k),\pi(k)) < x(0)^\top P_x x(0).
	\end{equation*}
\end{theorem}

Theorem \ref{thm:strict_performance_bound} states that one can expect a strict performance improvement if there are enough tokens in the bucket to support one extra transmission compared to the baseline periodic schedule. The result relies on a proof by contradiction, recognizing that if this extra transmission would not result in a strict performance improvement, then allowing arbitrarily many more transmissions would not result in a strict performance improvement either. This pathological case can be excluded by checking a condition stated in \cite[Theorem 4]{Wildhagen19_2}, which is omitted here for the sake of readability.

We demonstrate the performance results Theorem \ref{thm:performance_bound} and \ref{thm:strict_performance_bound} with the linearized batch reactor from \cite{Walsh01}, which allows to demonstrate two peculiar effects that might arise in certain combinations of system dynamics, token bucket parameters and prediction horizon: First, that the performance bound in Theorem \ref{thm:performance_bound} can be attained exactly and second, that the closed-loop transmission pattern of rollout ETC can degenerate to a periodic one. With a sampling time of \SI{0.1}{\second}, we have
\begin{equation*}
\begingroup % keep the change local
\setlength\arraycolsep{2pt}
{\small A = \begin{bmatrix}
1.178 &	0.001 &	0.512 &	-0.403 \\
-0.051 &	0.662 & -0.011 &	0.061 \\
0.076 &	0.335 &	0.561 &	0.382 \\
-0.001 &	0.335 &	0.089 &	0.850
\end{bmatrix}\hspace{-3pt}, \:
B = \begin{bmatrix}
0.004	& -0.0880 \\
0.467 &	0.001 \\
0.213 &	-0.235 \\
0.213 &	-0.016
\end{bmatrix}\hspace{-3pt}.}
\endgroup
\end{equation*}
The initial conditions are given by $x(0)=[1 \: 0 \: 1 \: 0]^\top$ and $w(0)=[ 0 \: 0]^\top$ and the plant is unconstrained, i.e., $\mathbb{X}\times\U=\R^{n+m}$. A token bucket TS with the parameters $g=3$, $c=8$ and $b=22$ has been assigned to the control loop, such that the base period is $M=3$ and the bandwidth limit is $\frac{g}{c}=\frac{3}{8}$. We use Variant 1 and two different prediction horizons $\overline{N}=3,6$.

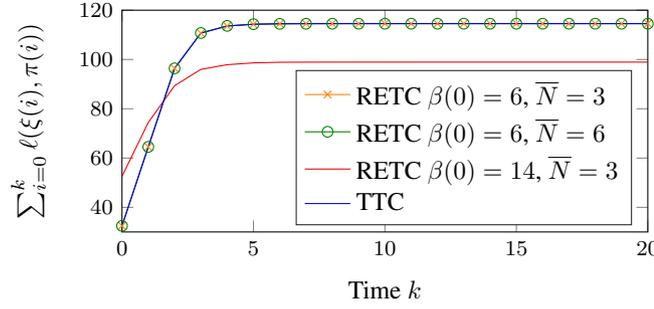
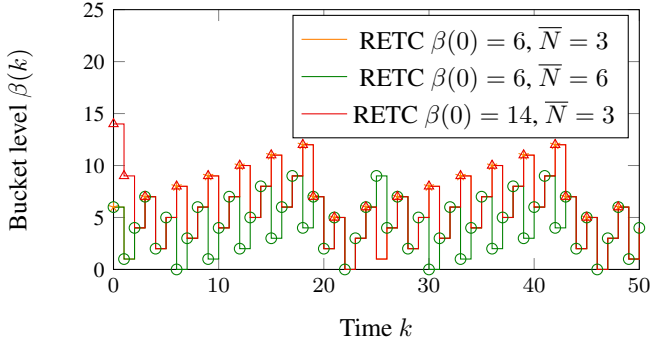
\begin{figure}[h]
	\begin{subfigure}{0.5\textwidth}
		\centering
		\begin{tikzpicture}

\begin{axis}[%
width=\columnwidth,
height=4.5cm,
xmin=0,
xmax=20,
xlabel={Time $k$},
ymin=30,
ymax=120,
ylabel={$\sum_{i=0}^{k} \ell(\xi(i),\pi(i))$},
axis background/.style={fill=white},
legend style={at={(0.97,0.03)}, anchor=south east, legend cell align=left, align=left, draw=white!15!black}
]

\addplot [color=istorange,mark=x]
table[row sep=crcr]{%
	0	32.4814264517711\\
	1	64.5382544564048\\
	2	96.4033694573245\\
	3	110.69957778419\\
	4	113.567524612028\\
	5	114.292422991806\\
	6	114.504299555297\\
	7	114.548148300781\\
	8	114.558531284494\\
	9	114.561088352871\\
	10	114.561821626216\\
	11	114.562096993248\\
	12	114.562211313163\\
	13	114.562247342469\\
	14	114.562259668622\\
	15	114.562263681708\\
	16	114.562264933953\\
	17	114.562265400753\\
	18	114.562265567288\\
	19	114.562265608042\\
	20	114.562265618949\\
	21	114.562265622102\\
	22	114.562265622879\\
	23	114.562265623136\\
	24	114.562265623216\\
	25	114.562265623232\\
	26	114.562265623237\\
	27	114.562265623238\\
	28	114.562265623238\\
	29	114.562265623238\\
	30	114.562265623238\\
	31	114.562265623238\\
	32	114.562265623238\\
	33	114.562265623238\\
	34	114.562265623238\\
	35	114.562265623238\\
	36	114.562265623238\\
	37	114.562265623238\\
	38	114.562265623238\\
	39	114.562265623238\\
	40	114.562265623238\\
	41	114.562265623238\\
	42	114.562265623238\\
	43	114.562265623238\\
	44	114.562265623238\\
	45	114.562265623238\\
	46	114.562265623238\\
	47	114.562265623238\\
	48	114.562265623238\\
	49	114.562265623238\\
	50	114.562265623238\\
};
\addlegendentry{RETC $\beta(0)=6$, $\overline{N}=3$}

\addplot [color=istgreen, mark=o]
table[row sep=crcr]{%
	0	32.4814264194991\\
	1	64.5382544000885\\
	2	96.4033693828326\\
	3	110.725056133103\\
	4	113.639030553032\\
	5	114.310558792527\\
	6	114.455576915913\\
	7	114.498101180952\\
	8	114.512398687022\\
	9	114.517331466549\\
	10	114.519349456264\\
	11	114.520076923467\\
	12	114.520331587188\\
	13	114.520427518919\\
	14	114.520458642636\\
	15	114.520469061899\\
	16	114.520472816551\\
	17	114.520474002234\\
	18	114.520474398459\\
	19	114.520474510449\\
	20	114.520474540231\\
};
\addlegendentry{RETC $\beta(0)=6$, $\overline{N}=6$}

\addplot [color=istred]
table[row sep=crcr]{%
	0	52.5659532590277\\
	1	74.418785296588\\
	2	89.3650060945971\\
	3	95.9838417604686\\
	4	97.8859346881137\\
	5	98.6190868068525\\
	6	98.8651443105046\\
	7	98.914548384606\\
	8	98.9271370573053\\
	9	98.9307740017239\\
	10	98.9316153176184\\
	11	98.931853445068\\
	12	98.9319310509788\\
	13	98.9319572636013\\
	14	98.9319670100044\\
	15	98.931970555639\\
	16	98.9319716952672\\
	17	98.9319721171542\\
	18	98.9319722640976\\
	19	98.9319723033605\\
	20	98.9319723142527\\
	21	98.9319723174701\\
	22	98.9319723182905\\
	23	98.9319723185718\\
	24	98.9319723186623\\
	25	98.9319723186804\\
	26	98.9319723186849\\
	27	98.9319723186861\\
	28	98.9319723186864\\
	29	98.9319723186865\\
	30	98.9319723186866\\
	31	98.9319723186866\\
	32	98.9319723186866\\
	33	98.9319723186866\\
	34	98.9319723186866\\
	35	98.9319723186866\\
	36	98.9319723186866\\
	37	98.9319723186866\\
	38	98.9319723186866\\
	39	98.9319723186866\\
	40	98.9319723186866\\
	41	98.9319723186866\\
	42	98.9319723186866\\
	43	98.9319723186866\\
	44	98.9319723186866\\
	45	98.9319723186866\\
	46	98.9319723186866\\
	47	98.9319723186866\\
	48	98.9319723186866\\
	49	98.9319723186866\\
	50	98.9319723186866\\
};
\addlegendentry{RETC $\beta(0)=14$, $\overline{N}=3$}

\addplot [color=istblue]
table[row sep=crcr]{%
	0	32.4814264517711\\
	1	64.5382544564048\\
	2	96.4033694573245\\
	3	110.69957778419\\
	4	113.567524612028\\
	5	114.292422991806\\
	6	114.504299555297\\
	7	114.548148300781\\
	8	114.558531284494\\
	9	114.561088352871\\
	10	114.561821626216\\
	11	114.562096993248\\
	12	114.562211313163\\
	13	114.562247342469\\
	14	114.562259668622\\
	15	114.562263681708\\
	16	114.562264933953\\
	17	114.562265400753\\
	18	114.562265565469\\
	19	114.56226560636\\
	20	114.562265619495\\
	21	114.562265623662\\
	22	114.56226562454\\
	23	114.562265624765\\
	24	114.562265624825\\
	25	114.562265624838\\
	26	114.562265624842\\
	27	114.562265624844\\
	28	114.562265624844\\
	29	114.562265624844\\
	30	114.562265624844\\
	31	114.562265624845\\
	32	114.562265624845\\
	33	114.562265624845\\
	34	114.562265624845\\
	35	114.562265624845\\
	36	114.562265624845\\
	37	114.562265624845\\
	38	114.562265624845\\
	39	114.562265624845\\
	40	114.562265624845\\
	41	114.562265624845\\
	42	114.562265624845\\
	43	114.562265624845\\
	44	114.562265624845\\
	45	114.562265624845\\
	46	114.562265624845\\
	47	114.562265624845\\
	48	114.562265624845\\
	49	114.562265624845\\
	50	114.562265624845\\
};
\addlegendentry{TTC}
\end{axis}
\end{tikzpicture}
		\vspace{-13pt}
		\caption{Cumulated stage cost up to time $k$.}
		\label{fig:cost_comparison}
	\end{subfigure}
	\begin{subfigure}{0.5\textwidth}
		\centering
\begin{tikzpicture}

\begin{axis}[%
width=\columnwidth,
height=5cm,
xmin=0,
xmax=50,
xlabel={Time $k$},
ymin=0,
ymax=25,
ytick={0,5,10,15,20,25},
yticklabels={0,5,10,15,20,25},
ylabel={Bucket level $\beta(k)$},
axis background/.style={fill=white},
]
\addplot[const plot, color=istorange] table[row sep=crcr] {%
	0	6\\
	1	1\\
	2	4\\
	3	7\\
	4	2\\
	5	5\\
	6	8\\
	7	3\\
	8	6\\
	9	9\\
	10	4\\
	11	7\\
	12	10\\
	13	5\\
	14	8\\
	15	11\\
	16	6\\
	17	9\\
	18	12\\
	19	7\\
	20	2\\
	21	5\\
	22	0\\
	23	3\\
	24	6\\
	25	1\\
	26	4\\
	27	7\\
	28	2\\
	29	5\\
	30	8\\
	31	3\\
	32	6\\
	33	9\\
	34	4\\
	35	7\\
	36	10\\
	37	5\\
	38	8\\
	39	11\\
	40	6\\
	41	9\\
	42	12\\
	43	7\\
	44	2\\
	45	5\\
	46	0\\
	47	3\\
	48	6\\
	49	1\\
	50	4\\
};
\addlegendentry{RETC $\beta(0)=6$, $\overline{N}=3$}

\addplot[const plot, color=istgreen] table[row sep=crcr] {%
	0	6\\
	1	1\\
	2	4\\
	3	7\\
	4	2\\
	5	5\\
	6	0\\
	7	3\\
	8	6\\
	9	1\\
	10	4\\
	11	7\\
	12	2\\
	13	5\\
	14	8\\
	15	3\\
	16	6\\
	17	9\\
	18	4\\
	19	7\\
	20	2\\
	21	5\\
	22	0\\
	23	3\\
	24	6\\
	25	9\\
	26	4\\
	27	7\\
	28	2\\
	29	5\\
	30	0\\
	31	3\\
	32	6\\
	33	1\\
	34	4\\
	35	7\\
	36	2\\
	37	5\\
	38	8\\
	39	3\\
	40	6\\
	41	9\\
	42	4\\
	43	7\\
	44	2\\
	45	5\\
	46	0\\
	47	3\\
	48	6\\
	49	1\\
	50	4\\
};
\addlegendentry{RETC $\beta(0)=6$, $\overline{N}=6$}

\addplot[const plot, color=istred] table[row sep=crcr] {%
	0	14\\
	1	9\\
	2	4\\
	3	7\\
	4	2\\
	5	5\\
	6	8\\
	7	3\\
	8	6\\
	9	9\\
	10	4\\
	11	7\\
	12	10\\
	13	5\\
	14	8\\
	15	11\\
	16	6\\
	17	9\\
	18	12\\
	19	7\\
	20	2\\
	21	5\\
	22	0\\
	23	3\\
	24	6\\
	25	1\\
	26	4\\
	27	7\\
	28	2\\
	29	5\\
	30	8\\
	31	3\\
	32	6\\
	33	9\\
	34	4\\
	35	7\\
	36	10\\
	37	5\\
	38	8\\
	39	11\\
	40	6\\
	41	9\\
	42	12\\
	43	7\\
	44	2\\
	45	5\\
	46	0\\
	47	3\\
	48	6\\
	49	1\\
	50	4\\
};
\addlegendentry{RETC $\beta(0)=14$, $\overline{N}=3$}

\addplot[const plot, color=istorange, draw=none, mark=star, mark options={solid, istorange}] table[row sep=crcr] {%
	0	6\\
	3	7\\
	6	8\\
	9	9\\
	12	10\\
	15	11\\
	18	12\\
	19	7\\
	21	5\\
	24	6\\
	27	7\\
	30	8\\
	33	9\\
	36	10\\
	39	11\\
	42	12\\
	43	7\\
	45	5\\
	48	6\\
};

\addplot[const plot, color=istorange, draw=none, mark=o, mark options={solid, istgreen}] table[row sep=crcr] {%
	0	6\\
	1	1\\
	2	4\\
	3	7\\
	4	2\\
	5	5\\
	6	0\\
	7	3\\
	8	6\\
	9	1\\
	10	4\\
	11	7\\
	12	2\\
	13	5\\
	14	8\\
	15	3\\
	16	6\\
	17	9\\
	18	4\\
	19	7\\
	20	2\\
	21	5\\
	22	0\\
	23	3\\
	24	6\\
	25	9\\
	26	4\\
	27	7\\
	28	2\\
	29	5\\
	30	0\\
	31	3\\
	32	6\\
	33	1\\
	34	4\\
	35	7\\
	36	2\\
	37	5\\
	38	8\\
	39	3\\
	40	6\\
	41	9\\
	42	4\\
	43	7\\
	44	2\\
	45	5\\
	46	0\\
	47	3\\
	48	6\\
	49	1\\
	50	4\\
};

\addplot[const plot, color=istgreen, draw=none, mark=triangle, mark options={solid, istred}] table[row sep=crcr] {%
	0	14\\
	1	9\\
	3	7\\
	6	8\\
	9	9\\
	12	10\\
	15	11\\
	18	12\\
	19	7\\
	21	5\\
	24	6\\
	27	7\\
	30	8\\
	33	9\\
	36	10\\
	39	11\\
	42	12\\
	43	7\\
	45	5\\
	48	6\\
};

\end{axis}
\end{tikzpicture}%
		\caption{Evolution of bucket level and transmission time points (indicated by marks).}
		\label{fig:bucket_level}
	\end{subfigure}
	\caption{Cost and bucket level for TTC and rollout ETC with different initial bucket levels and prediction horizons.}
\end{figure}

Figure \ref{fig:cost_comparison} shows the cumulated stage cost up to time $k$, i.e., $\sum_{i=0}^{k} \ell(\xi(i),\pi(i))$, for the optimal TTC and for rollout ETC with two different initial bucket levels, namely $\beta(0)=c-g +1= 6$ and $\beta(0)=2c-g+1 = 14$. Theorem \ref{thm:convergence} guarantees that the plant state and input converge to zero in all cases, which can also be seen from the fact that the cumulated stage costs converge to a certain level. In the case $\beta(0)=6$ and $\overline{N}=3$, we can see that the cost is exactly that of the optimal TTC. From Figure \ref{fig:bucket_level}, we can see why: the optimized transmission time points are periodic. We note that this peculiar case is allowed by Theorem \ref{thm:performance_bound}. This is different from the numerical example in the previous subsection, where rollout ETC achieved a strict performance improvement although also there, the overall number of transmissions was the same as for TTC. The periodic transmission pattern also vanishes when using a different prediction horizon, e.g., $\overline{N}=6$, as can be seen in Figure \ref{fig:bucket_level}. In contrast, if $\beta(0)=14$ and $\overline{N}=3$, the cost is strictly smaller for rollout ETC as expected from Theorem \ref{thm:strict_performance_bound}. We can see that also in this case, there is a transmission triggered as soon as there are enough tokens available. This suggests that for the particular prediction horizon $\overline{N}=3$, tokens are typically used up in a greedy manner by the RETC in this numerical example.
 
\subsection{Flexibility} \label{sec:flexibility}

In the numerical example in the previous subsection, we have observed that rollout ETC tends to trigger transmissions whenever possible and thereby draining the bucket, although the achieved gain in control performance might be very small. This is because the performance functional \eqref{eq:MPC_functional} considers only the plant state and input, whereas the bucket level is not represented. However, such a sampling behavior is not always favorable in practice. If there is, for instance, a set point change or a disturbance acting on the plant, then there would only be few tokens available to be able to react quickly to these unforeseen operating conditions. A promising strategy to counter this issue is to incentivize rollout ETC to trigger fewer transmissions than possible when the control loop has (almost) converged. Doing so would not much deteriorate control performance and the bucket would be filled in these phases. Then, the full bucket could be used to trigger many transmissions at times when this would indeed significantly improve control performance.

Such a behavior can be achieved if a small terminal cost on the bucket level is added, e.g., \begin{equation} \label{eq:terminal_cost_mod}
\tilde{V}_f(\xi,k)=V_f(\xi,k) + \sigma(b^2-\beta^2),
\end{equation}
where $\sigma>0$ is a small constant, and the terminal cost $V_f$ in \eqref{eq:MPC_functional} is replaced by $\tilde{V}_f$. The reasoning behind this strategy is that the transient behavior of the control loop is barely changed since the process costs $\ell$ and $V_f$ largely dominate the bucket cost $\sigma(b^2-\beta^2)$ when away from the set point. However, when the control system has almost converged, the terminal bucket cost takes effect and to minimize the cost functional, it is beneficial to trigger fewer transmissions than possible such that the bucket refills. The readers might ask themselves why $\beta$ is not included in the stage cost. This is because in general, convergence of the plant would be lost with this approach (see \cite[Remark 4]{jaumann2020saving}).

Under certain conditions, one can even guarantee for Variant 1 that the filling level converges to the upper sector of the bucket if the modified terminal cost $\tilde{V}_f$ is used.

\begin{theorem} \label{thm:flexibility}
	(\cite[Theorem 1]{jaumann2020saving}) Suppose Algorithm \ref{algo:rollout} is applied with Variant 1, the conditions of Theorem \ref{thm:convergence} hold and the modified terminal cost \eqref{eq:terminal_cost_mod} is used. Then, the same guarantees as in Theorem \ref{thm:convergence} hold. Moreover, if $\frac{c}{g}\notin\N$, $\beta(k)$ converges to $\N_{[\max(0,b-\overline{N}g),b]}$ as $k\to\infty$ in closed loop.
\end{theorem}

Theorem \ref{thm:flexibility} gives a convergence guarantee for the bucket level if $\frac{c}{g}$, i.e., the inverted sustainable rate, is not a natural number. The guaranteed set of convergence for $\beta(k)$ is dependent on the maximum prediction horizon $\overline{N}$, the bucket size $b$ and the refill rate $g$. For typical choices of these parameters, it effectively represents the upper sector of the bucket.

To demonstrate that a small terminal cost on the bucket level indeed achieves the desired behavior, we revisit the numerical example from Subsection \ref{sec:performance}. We use the modified terminal cost \eqref{eq:terminal_cost_mod} both for Variants 1 and 2 with $\sigma=10^{-6}$. The initial bucket level is $\beta(0)= b =22$.

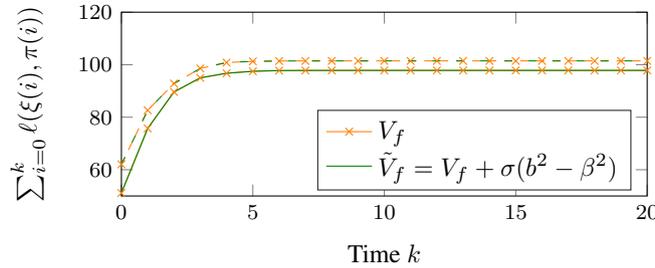
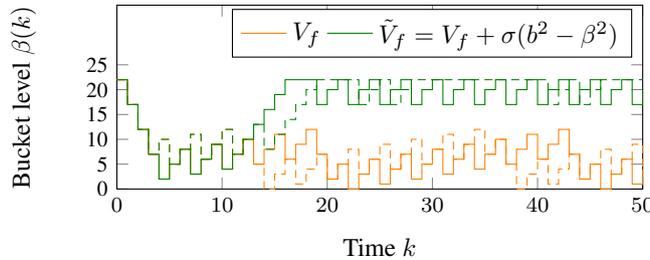
\begin{figure}
	\begin{subfigure}{0.5\textwidth}
		\centering
\begin{tikzpicture}

\begin{axis}[%
width=\columnwidth,
height=4cm,
xmin=0,
xmax=20,
xlabel={Time $k$},
ymin=50,
ymax=120,
ylabel={$\sum_{i=0}^{k} \ell(\xi(i),\pi(i))$},
axis background/.style={fill=white},
legend style={at={(0.97,0.03)}, anchor=south east, legend cell align=left, align=left, draw=white!15!black}
]

\addplot [color=istorange, mark=x]
table[row sep=crcr]{%
	0	51.1410892287537\\
	1	75.734021022121\\
	2	89.6287978501811\\
	3	94.999092105195\\
	4	96.7459182920112\\
	5	97.4984917895085\\
	6	97.7605612614473\\
	7	97.8129152574179\\
	8	97.8263765658605\\
	9	97.8303630015075\\
	10	97.8312614041628\\
	11	97.8315019389875\\
	12	97.8315751194736\\
	13	97.8316003638706\\
	14	97.8316100245159\\
	15	97.8316136601573\\
	16	97.8316148224688\\
	17	97.831615247801\\
	18	97.8316153941735\\
	19	97.8316154340054\\
	20	97.8316154451172\\
	21	97.831615448413\\
	22	97.8316154492581\\
	23	97.8316154495492\\
	24	97.8316154496432\\
	25	97.831615449662\\
	26	97.8316154496667\\
	27	97.831615449668\\
	28	97.8316154496683\\
	29	97.8316154496684\\
	30	97.8316154496684\\
	31	97.8316154496685\\
	32	97.8316154496685\\
	33	97.8316154496685\\
	34	97.8316154496685\\
	35	97.8316154496685\\
	36	97.8316154496685\\
	37	97.8316154496685\\
	38	97.8316154496685\\
	39	97.8316154496685\\
	40	97.8316154496685\\
	41	97.8316154496685\\
	42	97.8316154496685\\
	43	97.8316154496685\\
	44	97.8316154496685\\
	45	97.8316154496685\\
	46	97.8316154496685\\
	47	97.8316154496685\\
	48	97.8316154496685\\
	49	97.8316154496685\\
	50	97.8316154496685\\
};
\addlegendentry{$V_f$}

\addplot [color=istgreen]
table[row sep=crcr]{%
	0	51.1410892287537\\
	1	75.734021022121\\
	2	89.6287978501811\\
	3	94.999092105195\\
	4	96.7459182920112\\
	5	97.4984917895085\\
	6	97.7605612614473\\
	7	97.8129152574179\\
	8	97.8263765658605\\
	9	97.8303630015075\\
	10	97.8312614041628\\
	11	97.8315019389875\\
	12	97.8315748517728\\
	13	97.8315988782222\\
	14	97.8316078405197\\
	15	97.8316121663164\\
	16	97.8316151927111\\
	17	97.831618204027\\
	18	97.8316218377018\\
	19	97.8316236801885\\
	20	97.8316245917286\\
	21	97.8316250367641\\
	22	97.831625143092\\
	23	97.8316251682875\\
	24	97.8316251746197\\
	25	97.8316251766414\\
	26	97.8316251774141\\
	27	97.8316251777728\\
	28	97.8316251778597\\
	29	97.8316251778869\\
	30	97.831625177897\\
	31	97.8316251778991\\
	32	97.8316251778995\\
	33	97.8316251778996\\
	34	97.8316251778996\\
	35	97.8316251778996\\
	36	97.8316251778997\\
	37	97.8316251778997\\
	38	97.8316251778997\\
	39	97.8316251778997\\
	40	97.8316251778997\\
	41	97.8316251778997\\
	42	97.8316251778997\\
	43	97.8316251778997\\
	44	97.8316251778997\\
	45	97.8316251778997\\
	46	97.8316251778997\\
	47	97.8316251778997\\
	48	97.8316251778997\\
	49	97.8316251778997\\
	50	97.8316251778997\\
};
\addlegendentry{$\tilde{V}_f=V_f+\sigma(b^2-\beta^2)$}

\addplot [color=istorange, mark=x, dashed, mark options={solid}]
table[row sep=crcr]{%
	0	62.0513826203556\\
	1	82.6541772974914\\
	2	92.8517195376102\\
	3	98.5962403921424\\
	4	100.820940154968\\
	5	101.280277840904\\
	6	101.402678011688\\
	7	101.438356312026\\
	8	101.444834407109\\
	9	101.446422272263\\
	10	101.446964344098\\
	11	101.447127541152\\
	12	101.447173909451\\
	13	101.447185947702\\
	14	101.447189554406\\
	15	101.447190720993\\
	16	101.447191057173\\
	17	101.447191121287\\
	18	101.44719113329\\
	19	101.447191135476\\
	20	101.447191135866\\
	21	101.447191135946\\
	22	101.447191135965\\
	23	101.447191135969\\
	24	101.44719113597\\
	25	101.44719113597\\
	26	101.44719113597\\
	27	101.447191135971\\
	28	101.447191135971\\
	29	101.447191135971\\
	30	101.447191135971\\
	31	101.447191135971\\
	32	101.447191135971\\
	33	101.447191135971\\
	34	101.447191135971\\
	35	101.447191135971\\
	36	101.447191135971\\
	37	101.447191135971\\
	38	101.447191135971\\
	39	101.447191135971\\
	40	101.447191135971\\
	41	101.447191135971\\
	42	101.447191135971\\
	43	101.447191135971\\
	44	101.447191135971\\
	45	101.447191135971\\
	46	101.447191135971\\
	47	101.447191135971\\
	48	101.447191135971\\
	49	101.447191135971\\
	50	101.447191135971\\
};

\addplot [color=istgreen, loosely dashed]
table[row sep=crcr]{%
0	62.0513826203556\\
1	82.6541772974914\\
2	92.8517195376102\\
3	98.5962403921424\\
4	100.820940154968\\
5	101.280277840904\\
6	101.402678011688\\
7	101.438356312026\\
8	101.444834407109\\
9	101.446422272263\\
10	101.446964344098\\
11	101.447127541152\\
12	101.447174586408\\
13	101.447186893092\\
14	101.447190345881\\
15	101.447191315131\\
16	101.447191624594\\
17	101.447191820989\\
18	101.447192163425\\
19	101.447192921907\\
20	101.447194483862\\
21	101.447197430657\\
22	101.447204770446\\
23	101.447211095372\\
24	101.447216233867\\
25	101.447218124344\\
26	101.447218370513\\
27	101.447218415735\\
28	101.447218442599\\
29	101.44721846815\\
30	101.447218501299\\
31	101.447218559207\\
32	101.447218667749\\
33	101.447218862878\\
34	101.447219194317\\
35	101.447219731736\\
36	101.447220574071\\
37	101.447221863312\\
38	101.447223804802\\
39	101.447226697201\\
40	101.447232892888\\
41	101.447240219454\\
42	101.447246769236\\
43	101.447249970183\\
44	101.447250464196\\
45	101.447250544216\\
46	101.447250588856\\
47	101.447250703193\\
48	101.447251008013\\
49	101.447251671344\\
50	101.447252901971\\
};

\end{axis}
\end{tikzpicture}
		\vspace{-13pt}
		\caption{Cumulated stage cost up to time $k$ for Variant 1 (solid) and Variant 2 (dashed).}
		\label{fig:cost_comparison_2}
	\end{subfigure}
	\begin{subfigure}{0.5\textwidth}
		\centering
\begin{tikzpicture}

\begin{axis}[%
width=\columnwidth,
height=4cm,
xmin=0,
xmax=50,
xlabel={Time $k$},
ymin=0,
ymax=37,
ytick={0,5,10,15,20,25},
yticklabels={0,5,10,15,20,25},
ylabel={Bucket level $\beta(k)$},
axis background/.style={fill=white},
legend style={legend columns=2}
]

\addplot[const plot, color=istorange] table[row sep=crcr] {%
0	22\\
1	17\\
2	12\\
3	7\\
4	2\\
5	5\\
6	8\\
7	3\\
8	6\\
9	9\\
10	4\\
11	7\\
12	10\\
13	5\\
14	8\\
15	11\\
16	6\\
17	9\\
18	12\\
19	7\\
20	2\\
21	5\\
22	0\\
23	3\\
24	6\\
25	1\\
26	4\\
27	7\\
28	2\\
29	5\\
30	8\\
31	3\\
32	6\\
33	9\\
34	4\\
35	7\\
36	10\\
37	5\\
38	8\\
39	11\\
40	6\\
41	9\\
42	12\\
43	7\\
44	2\\
45	5\\
46	0\\
47	3\\
48	6\\
49	1\\
50	4\\
};
\addlegendentry{$V_f$}

\addplot[const plot, color=istgreen] table[row sep=crcr] {%
	0	22\\
	1	17\\
	2	12\\
	3	7\\
	4	2\\
	5	5\\
	6	8\\
	7	3\\
	8	6\\
	9	9\\
	10	4\\
	11	7\\
	12	10\\
	13	13\\
	14	16\\
	15	19\\
	16	22\\
	17	22\\
	18	22\\
	19	17\\
	20	20\\
	21	22\\
	22	17\\
	23	20\\
	24	22\\
	25	17\\
	26	20\\
	27	22\\
	28	17\\
	29	20\\
	30	22\\
	31	17\\
	32	20\\
	33	22\\
	34	17\\
	35	20\\
	36	22\\
	37	17\\
	38	20\\
	39	22\\
	40	17\\
	41	20\\
	42	22\\
	43	17\\
	44	20\\
	45	22\\
	46	17\\
	47	20\\
	48	22\\
	49	17\\
	50	20\\
};
\addlegendentry{$\tilde{V}_f=V_f+\sigma(b^2-\beta^2)$}

\addplot[const plot, color=istorange, densely dashed] table[row sep=crcr] {%
	0	22\\
	1	17\\
	2	12\\
	3	7\\
	4	10\\
	5	5\\
	6	8\\
	7	11\\
	8	6\\
	9	9\\
	10	12\\
	11	7\\
	12	10\\
	13	5\\
	14	0\\
	15	3\\
	16	6\\
	17	1\\
	18	4\\
	19	7\\
	20	2\\
	21	5\\
	22	8\\
	23	3\\
	24	6\\
	25	9\\
	26	4\\
	27	7\\
	28	10\\
	29	5\\
	30	8\\
	31	11\\
	32	6\\
	33	9\\
	34	12\\
	35	7\\
	36	10\\
	37	5\\
	38	0\\
	39	3\\
	40	6\\
	41	1\\
	42	4\\
	43	7\\
	44	2\\
	45	5\\
	46	8\\
	47	3\\
	48	6\\
	49	9\\
	50	4\\
};

\addplot[const plot, color=istgreen,densely dashed] table[row sep=crcr] {%
	0	22\\
	1	17\\
	2	12\\
	3	7\\
	4	10\\
	5	5\\
	6	8\\
	7	11\\
	8	6\\
	9	9\\
	10	12\\
	11	7\\
	12	10\\
	13	13\\
	14	8\\
	15	11\\
	16	14\\
	17	17\\
	18	20\\
	19	22\\
	20	22\\
	21	22\\
	22	22\\
	23	17\\
	24	20\\
	25	22\\
	26	17\\
	27	20\\
	28	22\\
	29	22\\
	30	22\\
	31	22\\
	32	22\\
	33	22\\
	34	22\\
	35	22\\
	36	22\\
	37	22\\
	38	22\\
	39	22\\
	40	22\\
	41	17\\
	42	20\\
	43	22\\
	44	17\\
	45	20\\
	46	22\\
	47	22\\
	48	22\\
	49	22\\
	50	22\\
};

\end{axis}

\end{tikzpicture}%
		\caption{Evolution of bucket level for Variant 1 (solid) and Variant 2 (dashed).}
		\label{fig:bucket_level_2}
	\end{subfigure}
	\caption{Cost and bucket level for rollout ETC with and without terminal cost on bucket level.}
\end{figure}

Figures \ref{fig:cost_comparison_2} and \ref{fig:bucket_level_2} depict the cumulated stage costs as well as the bucket levels, with and without the additional terminal cost on the bucket level and for both variants. From the two figures, we can see that our reasoning was correct. Starting with an initially full bucket, the rollout ETCs with and without terminal cost on the bucket level show almost the same transient behavior, as can be seen from the indistinguishable evolution of the cost. Only after about 10 time steps, when the plant has converged, the very small cost on the bucket level dominates over the process cost such that the bucket fills up to the upper sector when using $\tilde{V}_f$. This behavior can be observed for both variants. Since the bucket level has now converged to the upper sector of the bucket, there are once again many transmissions available in order to quickly react to future set point changes. As seen in the numerical example in Subsection \ref{sec:performance}, this will lead to a higher control performance for these subsequent set point changes and thus also in the long run.

\section{Conclusion and outlook} \label{sec:conclusion}

In this article, we provided an overview over recent research in rollout ETC, a control and transmission scheduling method aimed at reconciling classical ETC and TTC. In particular, we showed that by combining rollout ETC with the token bucket TS, it is both possible to satisfy a TS and thereby avoid congestion of the network (as in TTC), and at the same time to provide a flexible transmission triggering (as in ETC), and thereby to combine their advantages. A drawback of rollout ETC is that it is an optimization-based method, thereby potentially resulting in a higher computational complexity.

We presented two variants of rollout ETC under a common framework: one where the prediction horizon in the underlying OCP is cyclically shrinking, and one where the terminal ingredients are time-varying. We asserted that the prediction horizon and thus the computational complexity of Variant 2 is independent of the TS parameters, whereas Variant 1 requires a prediction horizon of at least the TS's base period. In exchange, Variant 1 offers theoretical guarantees on performance and convergence of the bucket level.

As rollout ETC is a rather new concept for control and transmission scheduling, there are still many open issues. For instance, output feedback and robustness to disturbances are important matters that are subject to our current research. Further, resilience to delays, dropouts and quantization, as well as extending the setup to more complex NCSs architectures and new types of TSs are largely open topics in rollout ETC.

\begin{acknowledgement}
  Funded by Deutsche Forschungsgemeinschaft (DFG, German Research Foundation) under Germany's Excellence Strategy - EXC 2075 - 390740016 and under grant AL 316/13-2 - 285825138.
\end{acknowledgement}

\bibliographystyle{plain}
\bibliography{bib_Rollout}

\newpage

\begin{wrapfigure}{l}{25mm} 
	\includegraphics[width=1in,height=1.25in,clip,keepaspectratio]{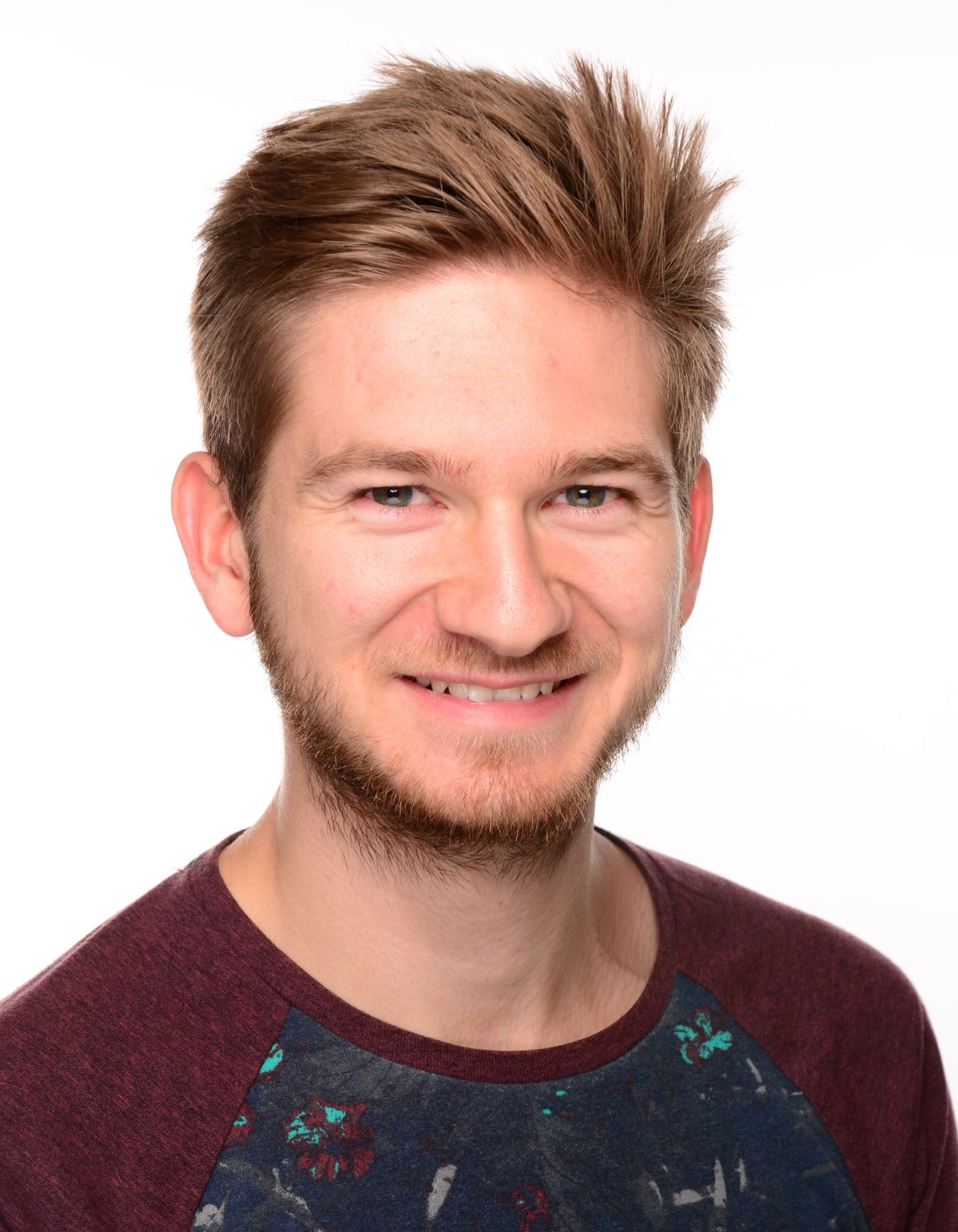}
\end{wrapfigure}\par
\textbf{Stefan Wildhagen} 
received the Master’s degree in Engineering Cybernetics from the University of Stuttgart, Germany, in 2018. He has since been a doctoral student at the Institute for Systems	Theory and Automatic Control under supervision of Prof. Allg\"ower and a member of the Graduate School Simulation Technology at the University of Stuttgart. His research interests are in the area of Networked Control Systems, with a focus on optimization-based scheduling and control as well as on data-driven methods.\par
\vskip10pt

\begin{wrapfigure}{l}{25mm} 
	\includegraphics[width=1in,height=1.25in,clip,keepaspectratio]{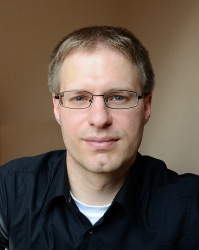}
\end{wrapfigure}\par
\textbf{Frank Dürr} 
is a senior researcher and lecturer at the Distributed Systems Department of the Institute
of Parallel and Distributed Systems (IPVS) at University of Stuttgart, Germany. He received
both his doctoral degree and diploma in computer science from University of Stuttgart. Frank Dürr
is currently leading the mobile computing and the software-defined networking (SDN) \& time-sensitive
networking (TSN) groups of the Distributed Systems Department. He has given tutorials on SDN at several national and international conferences, and as a lecturer he has been giving lectures and practical courses on networked systems and SDN. Besides SDN and TSN, Frank Dürr’s research interests
include mobile and pervasive computing, location privacy, and cloud computing aspects overlapping with these topics like mobile cloud and edge computing, or datacenter networks.\par
\vskip10pt

\begin{wrapfigure}{l}{25mm} 
	\includegraphics[width=1in,height=1.25in,clip,keepaspectratio]{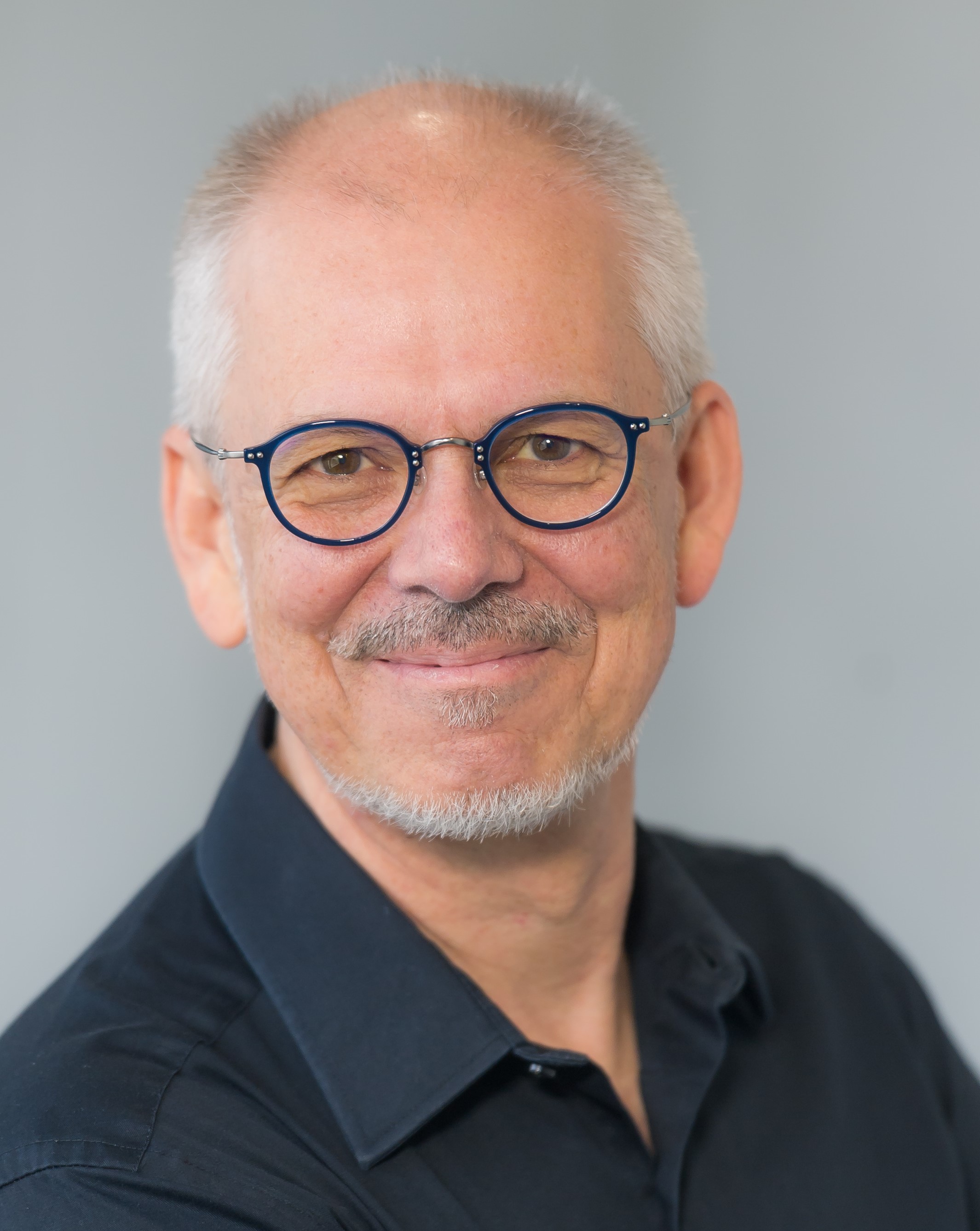}
\end{wrapfigure}\par
\textbf{Frank Allgöwer} 
is professor of mechanical engineering at the University of Stuttgart, Germany, and Director of the Institute for Systems Theory and Automatic Control (IST) there. \\
He is active in serving the community in several roles: Among others he has been President of the International Federation of Automatic Control (IFAC) for the years 2017-2020, Vicepresident for Technical Activities of the IEEE Control Systems Society for 2013/14, and Editor of the journal Automatica from 2001 until 2015. From 2012 until 2020 he served in addition as Vice-president for the German Research Foundation (DFG), which is Germany’s most important research funding organization. \\
His research interests include predictive control, data-based control, networked control, cooperative control, and nonlinear control with application	to a wide range of fields including systems biology.\par

\end{document}